             \documentclass[twocolumn,pra,showpacs]{revtex4} 
            \newcommand{\ads}{(a^{\dag}_{[l_{r},u_{r}]})^{s}}
            \newcommand{\adj}{a^{\dag}_{\a,j}}
            \newcommand{\adjp}{a^{\dag}_{\a^{\p},j^{\p}}}
            \newcommand{\aj}{a_{\a,j}}
            \newcommand{\ajp}{a_{\a^{\p},j^{\p}}}
            \newcommand{\ad}{a^{\dag}}
            \newcommand{\p}{\prime}
            
            \newcommand{\bdj}{b^{\dag}_{\b,j}}
            \newcommand{\bj}{b_{\b,j}}
            
            \newcommand{\bd}{b^{\dag}}
            \newcommand{\bdt}{(b^{\dag}_{[l_{i},u_{i}]})^{t}}
            
            \newcommand{\gs}{|\g s\rangle}
            
            \newcommand{\gsp}{|\g^{\p} s^{\p}\rangle}
            
            \newcommand{\gsdt}{|\g s,\d  t\rangle}
            \newcommand{\gsdtp}{|\g^{\p} s^{\p},\d^{\p} t^{\p}\rangle}
            \newcommand{\gsdtpp}{|\g^{\p\p}
            s^{\p\p},\d^{\p\p} t^{\p\p}\rangle}
            \renewcommand{\ap}{\alpha^{\p}}
            \renewcommand{\sp}{s^{\p}}
            
            \newcommand{\tp}{t^{\p}}
            \renewcommand{\a}{\alpha}
            \renewcommand{\b}{\beta}
            \newcommand{\cd}{c^{\dag}}
            \newcommand{\dd}{d^{\dag}}
            
            \newcommand{\g}{\gamma}
            \newcommand{\gp}{\gamma^{\p}}
            \renewcommand{\d}{\delta}
            \newcommand{\dep}{\delta^{\p}}
            \newcommand{\cg}{c_{\g,0}}
            \newcommand{\cdg}{c^{\dag}_{\g,0}}
            \newcommand{\dt}{|\d ,t\rangle}
            \newcommand{\dtp}{|\d^{\p},t^{\p}\rangle}
            \usepackage{graphics}
          \begin{document}
           \title{Fields of Iterated Quantum Reference
           Frames based on Gauge Transformations of Rational String States}
          \author{Paul Benioff\\
           Physics Division, Argonne National Laboratory \\
           Argonne, IL 60439 \\
           e-mail: pbenioff@anl.gov}
           \date{\today}

          \begin{abstract}
          This work is based on a description of quantum reference frames
          that seems more basic than others in the literature. Here
          a frame is based on a set of real and of complex numbers and
          a space time as a 4-tuple of the real numbers. There are many
          isomorphic frames as there are many isomorphic sets of real
          numbers. Each frame is suitable for construction of all physical
          theories as mathematical structures over the real and complex
          numbers. The organization of the frames into a field of frames is
          based on the representations of real  and complex numbers as
          Cauchy operators defined on complex rational states of
          finite qubit strings.

          The structure of the field is based on noting that the
          construction of real and complex numbers as Cauchy
          operators in a frame can be iterated to create new
          frames coming from a frame. Gauge transformations on the
          rational string states greatly expand the number of
          quantum frames as, for each gauge U, there is one frame
          coming from the original frame. Forward and backward iteration
          of the construction yields a two way infinite frame field with
          satisfying properties. There is no background space time and
          there are no real or complex numbers for the field as a whole.
          Instead these are relative concepts associated with each frame
          in the field. Extension to include qukit strings for different
          k bases, is described as is the problem of reconciling
          the frame field to the existence of just one frame with
          one background space time for the observable physical universe.
          \end{abstract}

           \maketitle

          \section{Introduction}
           As is well known, real and complex numbers are very
          important to physics.  All physical theories can be described
          as mathematical structures erected over the real and
          complex numbers. All theoretical predictions can be represented
          as real number solutions to equations given by the theories.  Also in
          many of the theories space time is represented
          as a $4$ tuple (or D-tuple in string theories) of the real numbers
          with an associated topology or metric structure.

          Another basic aspect of physics is that all physical representations of
          integers and rational numbers consist of a finite string of physical
          systems in states that correspond to a $k-ary$ representation of a number.
          Examples include the binary string representation
          of numbers used in most computations and computers, and
          the numerical output of measurements as finite strings of
          decimal digits. The string representation is also used in
          languages where words are strings of symbols in some alphabet.

          These aspects emphasize the importance of real numbers.
          Complex numbers are ordered pairs of real numbers
          with appropriate rules for addition and multiplication
          of the pair components.  A 4 tuple of real numbers is the
          representational basis for space time with topologies or
          metric structures added as appropriate. Also emphasized
          is the importance of finite strings of kits or qukits as $k-ary$
          representations of rational number approximations to real
          and complex numbers.  These will play a basic role here.

          Another aspect of central importance here is that
          there are many different reference frames in quantum
          theory \cite{Aharonov}. They play an important role  in quantum
          security protocols where a sender and receiver choose a
          reference frame for transmission and receipt of quantum
          information \cite{Bagan,Rudolph,Bartlett,vanEnk}.

          In this paper reference frames play a much more basic role
          than has been used so far.  In work done to date, such as
          that referenced above and \cite{Enk}, different reference
          frames are described within a fixed background space time.
          They are also based on  a fixed set of real and  complex
          numbers.  Here each reference frame is defined by or based
          on sets of real numbers, $R,$ and complex numbers, $C,$
          and $R^{4}$ as a basis of space time. The choice
          of topology and additional structure on $R^{4}$
          depends on the theory being considered in the frame.

          Since there are many different isomorphic
          representations of the real numbers there are many different
          isomorphic frames. Each frame is equivalent in that
          physical theories constructible by an observer in any one frame
          should have their equivalents in any other frame.
          This includes all theories that are mathematical structures
          over the real and complex numbers and describe the dynamics
          of systems in the background space time.

          If all these frames were not related and were
          independent of one another, this result would be of very
          limited interest.  However this is not the case.  The
          different frames are related. They form a structure that
          is a field of iterated quantum frames over the local
          and global gauge transformations of complex rational
          string states.

          The goal of this paper is to describe this structure and
          some of its properties. The structure is based on the
          quantum theory definitions of  real and complex numbers
          as (equivalence classes of) Cauchy operators. The definition
          of these operators is based on that for Cauchy sequences
          of real or complex rational string states \cite{BenRRCNQT}.
          These are states of finite qubit (or qukit) strings that
          are the quantum equivalent of the representation of real
          or complex rational numbers as finite digit strings in
          the binary (or $k-ary$) basis.

          The iteration of frames arises from the observation that
          in any frame, finite complex rational string states can be
          defined along with a set of Cauchy operators on these
          states. Since these Cauchy operators are real and complex
          numbers, they form the basis of another frame with a space
          time basis as a $4-tuple$ of the real number operators. This
          frame can be said to come from or emanate from the
          original frame. Since this process can be iterated one
          ends up with a finite, one way infinite, or two way
          infinite structure of iterated quantum frames.

          This one dimensional structure of iterated frames becomes
          two dimensional when one notes that there are a large
          number of representations of complex rational string
          states related by  global \cite{Enk} and local gauge
          transformations on the qubit string states. Extension
          of the structure to show the large number of different
          frames, one for each gauge transformation, coming from a frame,
          gives the resulting field structure of quantum frames.

          Most of this paper is devoted to describing and
          understanding various aspects of this structure.  The next
          section gives a representation of complex binary rational
          string states in terms of products of qubit creation
          operators acting on a vacuum state.  Basic arithmetic
          relations and operations are described. Section
          \ref{RRCNQT} repeats the definition in \cite{BenRRCNQT} of
          Cauchy sequences of complex binary rational string states
          and uses it to define operators that satisfy a
          corresponding Cauchy condition.  The real and
          complex number status of sets of Cauchy
          operators is based on noting that the proofs given in
          \cite{BenRRCNQT} can be applied here.

          Gauge transformations, as products of unitary operators in
          $SU(2),$ are introduced in Section \ref{GTR}.
          Transformations of rational string states and of the basic
          arithmetical relations and operations are described.
          Transformed Cauchy operators and the transformed Cauchy
          condition that they satisfy are also described.

          Section \ref{FIQF} uses these results to describe the
          basic structure of frames emanating from frames and the
          frame fields. The relation
          between observers in adjacent frames is discussed.
          Section \ref{SNB} completes the structure description by
          expanding it to include number strings to any base. It is
          seen that in this case each qukit operator in a gauge
          transformation can be represented as a product of operators
          in $U(1)\times SU(2)\times\cdots\times SU(p_{m}).$
          Here $p_{m}$ is the $mth$ prime number.

          The next section addresses the fundamental question of
          the connection of the structure to physics. The main problem
          is how to reconcile the infinity of reference frames,
          each with its own numbers and space time arena, to the
          physical existence of just one space time arena for physical
          systems ranging from cosmological to microscopic. In essence
          the problem is that all frames in the frame field are isomorphic or
          unitarily equivalent. But they are not the same.

          The solution of this problem is left to future work.
          However, it is noted that this requires some sort of
          process of merging or collapsing the frames into one frame,
          either by direct steps or by some limiting process so that
          the frames all become the same frame in the limit.
          Possible approaches include  use of  quantum error correction
          code methods, particularly those of decoherence free subspaces
          \cite{Lidar}. An example of how this might work is
          briefly summarized.

          It is clear that much work remains to be done.  However,
          the frame field picture is quite
          satisfactory in many ways. For the two way infinite field
          there is no absolute concept of given or abstract real or
          complex numbers. As is the case with loop quantum gravity
          \cite{Smolin}, there is no background space time associated
          with the frame field as a whole. These are
          relative concepts only in that for an observer in a frame,
          the real and complex numbers are abstract and the frame
          space time is the background. However, they are different for
          different frames. Also the frame tree has a basic
          direction built in, that of frames coming from frames.
          This picture also provides some concrete steps towards the
          construction of a coherent theory of physics and
          mathematics together \cite{BenTCTPM}.

          Another advantage of this approach is that it may
          provide a new approach to resolution of outstanding
          physical problems such as the integration of gravity with
          quantum mechanics. The new approach is based on the
          possibility that what an observer in a frame  $F$ sees as
          quantum dynamics of qubit string systems in states that
          represent numbers, and the corresponding dynamics of Cauchy
          operators, may be interpreted by an observer in a frame coming
          from $F$ as properties of his space time.  This would avoid
          the problem of an observer having to reconcile quantum
          mechanical aspects of systems and gravity for the same space
          time background of his own frame. In this approach it is
          possible that quantum dynamics of some systems in one frame
          correspond to metric and topological  properties of space
          time in another frame.

          If one regards complex numbers as ordered pairs of real
          numbers and space time as based on a 4 tuple of real
          numbers, then real numbers can be considered as the basis
          of the whole frame field construction. However complex
          numbers are basic because physical theories are mathematical
          structures over the complex numbers. Also a space time arena
          is needed for any theory describing the dynamics of systems
          moving in space time. A theory such as loop quantum gravity
          \cite{Smolin,Rovelli}, which is space time background
          independent, can be in any of the frames by ignoring the
          space time part of the frame. Aspects of string
          theory that emphasize the emergent properties of space
          times and the connections to general relativity
          \cite{Horowitz} may be relevant.

          Here complex numbers will be built directly by working
          with complex rational number states of qubit strings
          described by two types of annihilation creation operators,
          one for real and one for imaginary rational number states.
          In this way both components of complex numbers will be built
          up together rather than first building real numbers and
          then describing complex numbers as ordered pairs of real numbers.

          There does not appear to have been much study of quantum
          theory representations of real or complex numbers.
          The first mention of representations of real numbers as
          Hermitian operators appeared in the 1980s in a study of
          Boolean valued models of ZF set theory \cite{Takeuti}
          and appears to have been studied briefly  at that time
          \cite{Davis,Gordon}. Other more recent work on quantum
          representations of real numbers includes
          that in \cite{BenRRCNQT} and \cite{Litvinov,Corbett,Tokuo}.
          Of special note is recent work that emphasizes the
          possible importance of different sets of real numbers
          based on Takeuti's work in a category theoretic setting \cite{Krol}.

          \section{Real and Complex Binary Rational String
          States}\label{RCBRSS}

          There are a great many ways to represent real, imaginary,
          and complex binary rational string numbers in quantum theory.
          Here a representation based on annihilation creation
          operators for qubits is used that is different from that in
          \cite{BenRRCNQT} and \cite{BenRCRNQM} in that both
          the $0s$ and $1s$ will be included.  This coincides with
          the usual descriptions of quantum information as $0-1$
          string states of qubit strings.

          Let $\aj$ and $\adj$ where $\a =0,1$ and
          $j=\cdots,-1,0,1,\cdots$ be annihilation creation (AC) operators for
          one type of system in state $\a$ at integer site $j.$ Similarly
          let $\bj$ and $\bdj$ be AC operators for another type of
          system in state $\b =0,1$ at site $j$. For  bosons or fermions the AC
          operators satisfy commutation or anticommutation relations
          respectively:\begin{equation}\label{acomm}
         \begin{array}{c}[\aj,\adjp]=\delta_{j,j^{\p}}\delta_{\a,\ap} \\
         \mbox{$[\adj,\adjp]=[\aj,\ajp]=0$}\end{array}\end{equation}or
         \begin{equation}\label{aacomm}
         \begin{array}{c}\{\aj,\adjp\}=\delta_{j,j^{\p}}\delta_{\a,\ap} \\
         \{\adj,\adjp\}=\{\aj,\ajp\}=0.\end{array}\end{equation}
         with similar relations for the $b$ operators.

         Real and imaginary binary string rational states
         are represented by products of finite numbers of $a$ system
         (real) and $b$ system (imaginary) creation operators
         acting on the vacuum state $|0\rangle.$ Here $|0\rangle$
         denotes the empty qubit string. Also present in the
         operator strings are one $c$ system and one $d$ system
         creation operator, $c^{\dag}_{\g,l}$ and $d^{\dag}_{\d,k}.$
         The labels  $\g,\d = +,-$ are the signs and $l$ and $k$ are the
         integer locations of the "binal" points for the real and
         imaginary components of the complex string rational states.

         In the following $l$ and $k$ will be restricted to the
         value $l=k=0.$ This results in a simpler discussion of the
         arithmetic aspects of the rational string states.  To this
         end let $s:[l_{r},u_{r}]\rightarrow \{0,1\}$ and
         $t:[l_{i},u_{i}]\rightarrow \{0,1\}$ be functions from the
         integer intervals $[l_{r},u_{r}],[l_{i},u_{i}]$ to
         $\{0,1\}.$  Here $l_{r},l_{i}\leq 0$ and $u_{r},u_{i}\geq
         0.$ Complex binary rational string states are represented
         by\begin{equation}\label{binranum} \begin{array}{l}
         c^{\dag}_{\g,0}\ad_{s(l_{r}),l_{r}}
          \ad_{s(l_{r}+1),l_{r}+1}\cdots\ad_{s(u_{r}),u_{r}}d^{\dag}_{\d,0}
          \bd_{t(l_{i}),l_{i}} \cdots\bd_{t(u_{i}),u_{i}}|0\rangle \\
          =c^{\dag}_{\g,0}\ads d^{\dag}_{\d,0}\bdt|0\rangle.\end{array}
         \end{equation} The number $0$ is represented by
         $\cd_{+,0}\ad_{0,0}|0\rangle.$ The righthand expression in
         Eq. \ref{binranum} is a short way to express
         the product of AC operators in the string state. Note that
         the $a,b,c,d$ operators all commute with one another.

         In what follows it is useful to restrict $s$ and $t$ to
         exclude leading and trailing strings of $0s.$ To this end
         $s(l_{r})\mbox{ and }s(u_{r})\neq 0$ if $l_{r}<0
         \mbox{ and }u_{r}>0$ and $t(l_{i})\mbox{ and }t(u_{i})\neq 0$
         if $l_{i}<0 \mbox{ and }u_{i}>0.$

         In the representation used here the integer subscripts of
         the $a$ and $b$ operators represent positive or negative
         distances from the "binal" point and not locations on an
         underlying space lattice.  Also the $a$ and $b$
         systems could be at the same space locations or at
         different locations. Here it does not matter which is used
         as the $a$ and $b$ systems are distinct.

         One advantage of both quantum and classical representations
         that include a "binal" point is that the numerical value of
         any rational number state is invariant under any  space or time
         translation of the representation.  Values of
         $l_{r},u_{r},l_{i},u_{i}$ are the same wherever the string
         is located and the binal point location is always set equal
         to $0$.

         Another aspect of the representation used here is that a
         complex rational state can be represented either as one
         state or as a pair of states.  As one state it is
         represented as a string of $a,b$ creation operators and one $c,d$
         operator  acting on the vacuum state.  As a pair of states
         it is represented by
         $\cd_{\g,0}\ads|0\rangle,\dd_{\d,0} \bdt|0\rangle$
         representing the real and imaginary rational string states.
          Note that location $0$ is occupied by both a $\pm$ sign
          and a qubit state. An example of $\cd_{\g,0}\ads|0\rangle$
          with $\g = +$ is $|1011+0101\rangle.$ Here
          $l_{r}=-4$ and $u_{r}=3$ and the $+$ sign is understood
          to be at the same location as the digit immediately to
          the left. The usual representation of $1011+0101$ is
          $+1011.0101.$

         In what follows it will be helpful to work with a simpler
         notation.  To this end complex  binary rational string
         states are represented by $|\g s,\d t\rangle$ where
        \begin{equation}\label{shorthand}
         |\g s,\d  t\rangle \equiv c^{\dag}_{\g,0}\ads
          d^{\dag}_{\d,0}\bdt|0\rangle. \end{equation} Pure real and
          pure imaginary states are represented by $|\g
          s\rangle$ and $|\d t\rangle.$

         An operator $\tilde{N}$ can be defined whose eigenvalues
         correspond to the values of the numbers one usually associates
         with the string states. $\tilde{N}$ is the sum of two commuting
         operators $\tilde{N}^{R}$ and $\tilde{N}^{I}$ corresponding
         to the real and imaginary components. Each operator is the
         product of two commuting operators $\tilde{N}^{R,I}_{sp}$
         and $\tilde{N}^{R,I}_{v}.$
         These give respectively the sign and binary power shift
         given by the "binal" point location, and the binary value.
         One has \begin{equation}\label{numval}
         \tilde{N}=\tilde{N}^{R}+\tilde{N}^{I}=
         \tilde{N}^{R}_{sg}\tilde{N}^{R}_{v}+\tilde{N}^{I}_{sg}
         \tilde{N}^{I}_{v}\end{equation} where
         \begin{equation}\label{NRIspv}\begin{array}{c}
         \tilde{N}^{R}_{sg}=\sum_{\g}\g \cdg\cg\\
         \tilde{N}^{I}_{sg}=\sum_{\d}\d d^{\dag}_{\d,0}d_{\d,0} \\
          \tilde{N}^{R}_{v}=\sum_{\a,j}\a 2^{j}\adj\aj\\
          \tilde{N}^{I}_{v}=\sum_{\b,j}i\b 2^{j}\bdj\bj\end{array}
         \end{equation}

         Note that the eigenvalues of these operators are
         independent of the ordering of the $a,b,c,d$ operators.
         For bosons the state is independent of the
         order of the operators, for fermions a fixed ordering,
         such as that shown in the equation with $j$
         increasing from right to left will be used here. This
         operator is defined for reference purposes only as it is
         not used in the following.

         Arithmetic properties and operations can be defined
         independent of $\tilde{N}$. It is sufficient to define the
         basic relations  as all properties are
         combinations of the basic ones using the logical
         connectives. The basic properties or relations are arithmetic equality
         $=_{A}$ and ordering $\leq_{A}.$ Let $\gs$ and $\gsp$ be two
         real rational  states and $\dt$ and $\dtp$ be two
         imaginary rational states. Then\begin{equation}\label{defequalA}
         \begin{array}{c}\gs =_{A}\gsp \mbox{ if and only if } \g =\g^{\p}
         \mbox{ and }s=s^{\p} \\ \dt =_{A} \dtp\mbox{ if and only if }
         \d =\d^{\p} \mbox{ and } t=t^{\p}.\end{array} \end{equation}
         The definition says that two real or imaginary rational string
         states are arithmetically equal if they have the same signs and
         distributions of $1s$ and $0s$ on the same integer intervals. Two complex
         rational states are equal if both the real and imaginary parts are equal.

         For any function $s$ define $1_{s}=\{j:s(j)=1\}$ as the set of
         sites on which $s(j)=1.$ Arithmetic ordering on positive real
         rational states is defined by
         \begin{equation}\label{deforderA}
         |+, s\rangle \leq_{A}|+,s^{\p}\rangle
         \mbox{ if } 1_{s}\leq 1_{s^{\p}} \end{equation}
         where\begin{equation}\label{deforderAone}
         \begin{array}{c}1_{s}< 1_{s^{\p}}\mbox{ if there is a $j$
         where $j$ is in $1_{s^{\p}}$ and not in
         $1_{s}$} \\ \mbox{and for all $k>j$ $k\epsilon 1_{s}$
         iff $k\epsilon 1_{s^{\p}}$}.\end{array}
         \end{equation}  The extension to zero and negative
         real rational states is given by\begin{equation}\label{0negordr}
         \begin{array}{c} \cd_{+,0}\ad_{0,0}|0\rangle
         \leq_{A} c^{\dag}_{+,0}\ads |0\rangle \mbox{ for all  $s$}\\
         |+,0,s\rangle \leq_{A}|+,0,s^{\p}\rangle
         \rightarrow |-,0,s^{\p}\rangle\leq_{A}
         |-,0,s\rangle.\end{array}\end{equation}
         Similar relations hold for the imaginary components of
         complex string states.

         Definitions of operators for the arithmetic operations
         $\tilde{+}_{A},\tilde{\times}_{A},\tilde{-}_{A},
         \tilde{\div}_{A,-\ell}$, addition,
         multiplication, subtraction, and division to any accuracy
         $2^{-\ell}$ can be obtained by converting those given in
         \cite{BenRCRNQM} to the definitions of complex rational string
         numbers used here. Detailed definitions in terms of AC
         operators will not be given here as nothing new is added.
         One should note that these operators are binary. For instance
         for addition one has\begin{equation}\label{addn}
         \tilde{+}_{A}\gsdt\times\gsdtp=\gsdt\times\gsdtpp.\end{equation}Here
         the state $\gsdtpp$ is the result of the addition in that
         \begin{equation}\label{resaddn}\gsdtpp=_{A}\gsdt+_{A}\gsdtp.
         \end{equation} The righthand term expresses addition in the
         usual way and will often be used here. Similar considerations
         hold for the other three arithmetic operators.

         The arithmetic relations also have a direct
         quantum theoretic expression in terms of AC operator
         products and sums. These expressions are obtained by
         translating the conditions expressed in the definitions
         into operator expressions  such that states are in the
         subspace of eigenvalue $1$ for a projection operator if
         and only if the statements are true.

         For example $\gs =_{A}\gsp$ is true if and only if
         \begin{equation}\label{PequalA}
        P_{=_{A}}\gs\times\gsp = \gs\times\gsp\end{equation}
        where \begin{equation}\label{QMequalA} P_{=_{A}}\gs\times\gsp=
         \sum_{\nu}\sum_{w}\prod_{j\epsilon
         w}P_{1,j}P_{\nu,0}\gs\times P_{1,j}P_{\nu,0}\gsp.
        \end{equation} Here $\nu=+,-$, and the $w$ sum is over all
        finite subsets of integers. The righthand projection
        operators are given by
        \begin{equation}\label{projAC} P_{1,j}=\ad_{1,j}a_{1,j}
        \;\;\;P_{\nu,0}=\cd_{\nu,0}c_{\nu,0}. \end{equation}

        The  projection operator for $\leq_{A},$ $P_{\leq_{A}},$ is
        given by $P_{\leq_{A}}=P_{=_{A}}+P_{<_{A}}.$  For positive
        states (those with $\g=\g^{\p}=+$),
        \begin{equation}\label{Plessth}\begin{array}{l}
        P_{<_{A}}|+,0,s\rangle \times |+,0,s^{\p}\rangle=
        \sum_{w,w^{\p}}\sum_{j\epsilon
        w}\prod_{k>j}^{max\{w,w^{\p}\}} \\ \hspace{1cm}
        \sum_{i=0,1}P_{0,j}P_{i,k}|+,0,s\rangle\times
        P_{1,j}P_{i,k}|+,0,s^{\p}\rangle.\end{array}
        \end{equation} Here $max\{w,w^{\p}\}$ is the
        largest number in both $w$ and $w^{\p}.$ Eq. \ref{Plessth}
        is set up to correspond to the expression in Eq.
        \ref{deforderAone}. For extension to negative states one has
        $P_{<_{A}}|-,0,s\rangle\times|-,0,s^{\p}\rangle =
        |-,0,s\rangle\times|-,0,s^{\p}\rangle$
        if and only if $P_{<_{A}}|-,0,s^{\p}\rangle
        \times|+,0,s\rangle = |-,0,s^{\p}\rangle\times
        |-,0,s\rangle.$

         It is to be emphasized that all arithmetic operations and
         relations are distinguished from quantum mechanical
         relations and operations by subscripts.  For instance
         $=_{A}$ and $+_{A}$ denote arithmetic equality and
         addition: $=$ and $+$ denote quantum mechanical state
         equality and linear superposition.

         \section{Representations of Real and Complex Numbers in
         Quantum Theory}\label{RRCNQT}
         Here the  binary complex rational string states are used
         to describe representations of real and complex numbers
         in quantum theory. The work makes use of the results
         in \cite{BenRRCNQT}. Real, imaginary, and complex
         numbers are represented  as Cauchy sequences of real,
         imaginary, and complex rational string states. The definitions,
         given in \cite{BenRRCNQT} for binary representations of rational
         numbers that suppress use of the $0s$, apply here also.

         Let $\{\gs_{n}\}=\{\gs_{n}:n=0,1,\cdots\}$ denote a
         sequence of states $|\g_{n}s_{n}\rangle$ where the
         possible $n$ dependence of the sign, and qubit states in a
         string $s_{n}$  defined on the interval $[l_{r,n},u_{r,n}]$
         is made explicit. This shows that all string state parameters can depend
         on the position $n$ in the sequence. The sequence satisfies
         the Cauchy condition if \begin{equation}\label{cauchyr}
          \begin{array}{c}\mbox{ For each $\ell$ there is an $h$
          where for all $j,k>h$} \\
          |(|\g_{j}s_{j}-\g_{k}s_{k}|)\rangle
          <_{A}|+,-\ell\rangle.\end{array} \end{equation} In this
          definition $|(|\g_{j}s_{j}-\g_{k}
          s_{k}|)\rangle\equiv ||\g_{j}s_{j}\rangle
          -_{A}|g_{k}s_{k}\rangle|_{A}$ is the state that is
          the arithmetic absolute value of the arithmetic difference
          between the states $|\g_{j},s_{j}\rangle$ and
          $|\g_{k},s_{k}\rangle.$ The Cauchy condition says that
          this state is arithmetically less than or equal to the
          state $|+,-\ell\rangle=\cd_{+,0}
          \ad_{0,-1}\ad_{0,-2}\cdots\ad_{0,-\ell+1}\ad_{1,-\ell}|0\rangle$
          for all $j,k$ greater than some $h$. The $\tilde{N}$
          eigenvalue of the state $|+,-\ell\rangle$ is $2^{-\ell}.$

          This definition can be easily extended to imaginary and
          complex binary rational string states. For complex states
          the Cauchy condition is \begin{equation}\label{cauchyri}
          \begin{array}{c}\mbox{ For each
          $\ell$ there is an $h$ where for all
          $j,k>h$} \\  |(|\g_{j}s_{j}-\g_{k}s_{k}|)\rangle
          <_{A}|+,-\ell\rangle\mbox{ and }\\
          |(|\d_{j}t_{j}-\d_{k}t_{k}|)\rangle
          <_{A}|+,-\ell\rangle\end{array}. \end{equation}

          It was also seen in \cite{BenRRCNQT} that the Cauchy condition
          can be extended to sequences of linear superpositions of
          complex rational states. Let $\psi_{n}
          =\sum_{\g,s}\sum_{\d,t}\gsdt\langle\g s,\d t|\psi_{n}\rangle.$
          Then\begin{equation}\label{Pjml}
          \begin{array}{l}P_{j,m,\ell}= \\ \hspace{0.5cm}\sum_{\g,s}\sum_{\d,t}
          \sum_{\gp,\sp}\sum_{\dep,\tp}
          |\langle\g s,\d t|\psi_{j}\rangle
          \langle\gp \sp,\dep,\tp|\psi_{m}\rangle|^{2} \\
          \hspace{2cm} (|(|\g s-\gp \sp|)\rangle \leq_{A}
          |+,-\ell\rangle \mbox{ and } \\
          \hspace{2cm} |(|\d t-\dep  \tp|)\rangle \leq_{A}
          |+,-\ell\rangle )\end{array}\end{equation} is the
          probability that the arithmetic absolute value
          of the arithmetic differences between the real parts
          and the imaginary parts of $\psi_{j}$ and $\psi_{k}$ are
          each arithmetically less than or equal to
          $|+,-\ell\rangle.$ The sequence $\{\psi_{n}\}$
          satisfies the Cauchy condition if $P_{\{\psi_{n}\}} =1$
          where\begin{equation}\label{limPjkl}
          P_{\{\psi_{n}\}} =\liminf_{\ell\rightarrow\infty}
          \limsup_{h\rightarrow\infty}\liminf_{j,k>h}P_{j,m,\ell}.
          \end{equation} Here $P_{\{\psi_{n}\}}$ is the probability
          that the sequence $\{\psi_{n}\}$ satisfies the Cauchy condition.

          An equivalence relation is defined between Cauchy
          sequences by noting that $\{\gsdt_{n}\}\equiv
          \{\gsdtp_{m}\}$ if the condition of Eq. \ref{cauchyri}
          is satisfied with $\gp_{k},\sp_{k}$ replacing
          $\g_{k},s_{k}$ and $\dep_{k}\tp_{k}$
          replacing $\d_{k}t_{k}.$  Similarly
          $\{\psi_{n}\}\equiv\{\psi^{\p}_{m}\}$ if
          $P_{\{\psi_{n}\}\equiv\{\psi^{\p}_{m}\}}=1$ where
          $P_{\{\psi_{n}\}\equiv\{\psi^{\p}_{m}\}}=1$ is given by
          Eqs. \ref{Pjml} and \ref{limPjkl} with $\psi^{\p}_{k}$
          replacing $\psi_{k}$ in Eq. \ref{Pjml}.

          The equivalence relation is used to collect Cauchy
          sequences into equivalence classes. The sets of
          equivalence classes of Cauchy sequences of real,
          imaginary, and complex rational string states and their
          linear superpositions are the real numbers $R_{A}$,
          imaginary numbers $I_{A}$ and complex numbers $C_{A}.$ It
          is shown in \cite{BenRRCNQT} that $R_{A},I_{A}, C_{A}$ satisfy the
          relevant axioms.

          The new step here is to replace sequences of binary
          rational string states with operators and require that
          the operators satisfy the Cauchy condition.
          To achieve this one needs to define the rational
          string states that are the nonnegative integers. The state $\gs =
          \cd_{\g,0}\ads|0\rangle$ is a nonnegative integer if
          $\g =+$ and $l_{r}=0.$  This corresponds to the observation that
          bit string numbers such as $ +10010.=10010+$ are nonnegative integers.

          Let $\tilde{O}$ be a linear operator whose domain is the
          Fock space over the nonnegative integer states
          $\{|+,s\rangle\}$ and whose range is  the Fock space
          over the binary complex rational string
          states.   One has \begin{equation}\label{defOgen}
          \tilde{O}|+,s^{\p}\rangle=\sum_{\g,s}\sum_{\d,t}
          \gsdt \langle \g s,\d t|\tilde{O}|+,s^{\p}\rangle.
          \end{equation} In general the state
          $\tilde{O}|+,s^{\p}\rangle$ is a linear superposition of
          binary complex rational states. If the $\tilde{O}$ matrix
          elements are nonzero only if $t=\underline{0}$ ($s= \underline{0}$)
          then $\tilde{O}|+,s^{\p}\rangle$ is a superposition of real
          (imaginary) rational string states.

          Here the interest is in operators that are normalized and
          satisfy the Cauchy condition. Normalization means that
          \begin{equation}\label{normO}\langle +,s|\tilde{O}^{\dag}
          \tilde{O}|+,s\rangle =1\end{equation}
          for all integer $s.$  For the Cauchy condition, first restrict
          $\tilde{O}$ so that $\tilde{O}|+,s^{\p}\rangle$ is just
          one binary rational state instead of a superposition of more than
          one of them.  Then $\tilde{O}$ satisfies the Cauchy
          condition if \begin{equation}\label{cauchyO}
          \begin{array}{c}\mbox{ For each
          $|+,-\ell\rangle$ there is a state $|+,s_{h}\rangle$}
          \\ \mbox{ where for all states $|+,s_{j}\rangle,
          |+,s_{k}\rangle >_{A}|+,s_{h}\rangle$} \\
           |P_{R}\tilde{O}|+,s_{j}\rangle
           -_{A}|P_{R}\tilde{O}|+,s_{k}\rangle |_{A}
          <_{A}|+,-\ell\rangle\mbox{ and }\\  |P_{I}\tilde{O}|+,s_{j}\rangle
           -_{A}|P_{I}\tilde{O}|+,s_{k}\rangle |_{A}
          <_{A}|+,-\ell\rangle \end{array}. \end{equation} Here
          $P_{R}$ and $P_{I}$ are projection operators onto
          the subspaces of real and imaginary rational string
          states and $>_{A}$ is the natural arithmetic ordering of
          the integer states.  This definition mirrors that of Eq.
          \ref{cauchyri} with integer states replacing $h,j,k.$
          If the states $\tilde{O}|+,s\rangle$ are linear superpositions as in Eq.
          \ref{defOgen},  then the Cauchy condition is given by Eq.
          \ref{Pjml} with $\psi_{j}$ and $\psi_{m}$ replaced by
          $\tilde{O}|+,s_{j}\rangle$ and
          $\tilde{O}|+,s_{m}\rangle.$

          From this one sees that the set of all operators
          $\tilde{O}$ that are normalized and are Cauchy (satisfy
          Eq. \ref{cauchyO} or \ref{Pjml} with the replacements
          indicated) can be gathered into equivalence classes of
          real, imaginary, or complex numbers.  In what follows
          elements of equivalence classes will be considered as
          representatives of the classes. Thus $\tilde{O}$ is a pure
          real or pure imaginary number if the respective imaginary or real
          parts of $\tilde{O}|+,s\rangle\rightarrow 0$ as
          $|+,s\rangle\rightarrow\infty.$ These will be referred to
          in the following as Cauchy operators.

          Let $\mathcal{R},\mathcal{I},\mathcal{C}$ denote the sets
          of Cauchy operators that are pure real, pure
          imaginary, or complex.  As is well known, both
          $\mathcal{R}$ and $\mathcal{I}$ are contained in $\mathcal{C}.$
          The proofs that the Cauchy sequences of rational string
          states have the relevant axiomatic properties,
          given in \cite{BenRRCNQT}, also apply here. The
          proofs are based on definitions of the basic relations and
          operations such as $=_{R};\leq_{R}$ and
          $+_{R},\times_{R},-_{R},\div_{R}$ that are based on
          corresponding definitions of these relations and
          operations for rational string states.

          For example, if $\tilde{O}$ and $\tilde{O}^{\p}$ are
          normalized Cauchy operators and, for each integer state $|s\rangle$,
          $\tilde{O}|s\rangle$ and
          $\tilde{O}^{\p}|s\rangle$  are single rational string states,
          then one can define  $\tilde{O^{\p\p}}$ for all $|s\rangle$ by
        \begin{equation}\label{OOpOpps}\tilde{O^{\p\p}}|s\rangle =
        \tilde{O}|s\rangle +_{A}\tilde{O}^{\p}|s\rangle.
        \end{equation} (From here on the $+$ sign will be
        suppressed in the nonnegative integer states.)
        This equation states that the
        left hand state is defined as the quantum state
        that is the arithmetic sum of the right hand states.
        Note how different this is from the usual quantum
        mechanical superposition.  For example, the equation has a very
        different meaning if $+_{A}$ is replaced by $+$.

        The usual numerical relation $|x+y-(x^{\p}+y^{\p})
        |\leq |x-x^{\p}|+|y-y^{\p}|$ applied here gives
        \begin{equation}\label{cauchtOOp}\begin{array}{l}
        |P_{R}\tilde{O}|s\rangle_{k}+_{A}P_{R}\tilde{O}^{\p}|s\rangle_{k}-_{A}
        (P_{R}\tilde{O}|s\rangle_{j}+_{A}P_{R}\tilde{O}^{\p}|s\rangle_{j})|_{A}\\
        \hspace{1cm}\leq_{A}|P_{R}\tilde{O}|s\rangle_{k}-_{A}
        P_{R}\tilde{O}|s\rangle_{j}|_{A}
        \\ \hspace{1cm} +_{A}|P_{R}\tilde{O}^{\p}|s\rangle_{k}-_{A}P_{R}
        \tilde{O}^{\p}|s\rangle_{j}|_{A}.\end{array}
        \end{equation} Since $\tilde{O}$ and
        $\tilde{O}^{\p}$ are Cauchy, it follows that $\tilde{O}^{\p\p}$ is
        Cauchy for the real component. Repeating the argument for
        the imaginary component shows that $\tilde{O}^{\p\p}$ is
        also Cauchy for the imaginary component.  One concludes then
        that $\tilde{O}^{\p\p}$ is the numerical sum of $\tilde{O}$
        and $\tilde{O}^{\p}$,\begin{equation}\label{OOpsum}
        \tilde{O}^{\p\p}=\tilde{O}+_{R}\tilde{O}^{\p}.
        \end{equation} Note that the equality is the operator
        equality in quantum mechanics but the sum is not; it is
        the numerical sum as defined for the addition of two complex
        numbers.  To coincide with standard usage the subscript $R$ is used
        for real, imaginary, and complex numbers  The presence of the
        subscript helps to avoid confusion with other uses of these symbols.

        If the states $\tilde{O}|s\rangle$ and
        $\tilde{O}^{\p}|s\rangle$ are linear superpositions as in Eq.
        \ref{defOgen} then Eq. \ref{OOpOpps} is not valid as
        the result of arithmetic addition is a density operator
        state\begin{equation}\label{Orho}
        \begin{array}{l}\rho_{\tilde{O}|s\rangle +\tilde{O}^{\p}|s\rangle}
        =\sum_{\g s,\d t}\sum_{\gp
        \sp,\dep t^{\p}}\sum_{\g^{\p\p}s^{\p\p},\d^{\p\p}t^{\p\p}}|\langle\g
         s,\d t|\tilde{O}|s\rangle|^{2} \\ \times \langle \gp \sp,\dep
        \tp |\tilde{O}^{\p}|s\rangle\langle
        s|\tilde{O}^{\p}|\g^{\p\p}s^{\p\p},\d^{\p\p}t^{\p\p}\rangle
        \\ \times (|\g s,\d t\rangle +_{A}|\gp \sp,\dep \tp\rangle)
        (\langle \g s\,\d t|+_{A}\langle \g^{\p\p}s^{\p\p},\d^{\p\p}
        t^{\p\p}|.\end{array}\end{equation} The proof that this is
        Cauchy uses the diagonal terms only $(\gp
        =\g^{\p\p},\sp=s^{\p\p},\dep =\d^{\p\p},\tp =t^{\p\p})$ and
        is given in \cite{BenRRCNQT}.

        The definition of $+_{R}$ applies in this case also because
        the definitions in Eqs. \ref{OOpOpps} and \ref{OOpsum} apply
        to equivalence classes containing the operators of Eq.
        \ref{OOpOpps}. Since every equivalence class contains
        operators that satisfy Eq. \ref{OOpOpps}, it applies to
        $\tilde{O},\tilde{O}^{\p}$ in Eq. \ref{Orho}.

        Similar definitions can be given for the other numerical
        operations, $\times_{R},-_{R},\div_{R}$ by converting those
        given in \cite{BenRRCNQT} to the operator definitions used
        here. These definitions also apply to more complex
        analytical operations  and functions such as $df(x)/dx$ and integrals
          $\int_{a}^{b}f(x,y)dx$ where $ a,b,x,y$ are operator
          variables ranging over $\mathcal{R},\mathcal{I},\mathcal{C}.$

          It follows that all of real and complex analysis and all
          mathematics used in physics that is based at some point on
          real and complex numbers can be based on the sets $\mathcal{R},
          \mathcal{I},\mathcal{C}$ of (equivalence classes of) these  operators.
          Also in any theory that models space time  as $R^{4},$
          or $R^{3}$ and $R$ nonrelativistically, can be based on
          space time modeled as ${\mathcal{R}}^{4},$ or as
          ${\mathcal{R}}^{3}$ and $\mathcal{R}.$  In this sense one
          sees that the complex rational states as defined here
          provide a reference frame \cite{Toller} for physical
          theories and for space time that is based on the real
          and complex numbers as normalized Cauchy operators for the
          complex rational states.

          This description extends to many different reference frames that
          are unitarily equivalent to the frame described above. To see this
          let $U$ be an arbitrary unitary transformation of the binary complex
          rational string states.  It is clear  that in general, arithmetic
          relations and operations  are not conserved under $U$.
          For example one can have $\gsdt =_{A}\gsdtp$ but
          $U\gsdt \neq_{A}U\gsdtp.$ Also the Cauchy
          property is not conserved for most $U$ in that if
          $\tilde{O}$ is Cauchy on $\{\gsdt\}$ it is not Cauchy on
          $\{U\gsdt\}.$

          All the desired properties can be recovered by replacing
          the original definitions of arithmetic relations and
          operations, described in the previous sections, by
          transformed relations and operations.  That is, $=_{A},\leq_{A}$
          and $+_{A},\times_{A},-_{A},\div_{A,\ell}$ are replaced by
          appropriately defined relations and operations
          $=_{AU},\leq_{AU}$ and $+_{AU},\times_{AU},-_{AU},\div_{AU,\ell},$
          and $\tilde{O}$ is replaced  by $\tilde{O}_{U}.$ Use of these
          replacements in the descriptions of the previous sections gives an
          entirely equivalent description of rational string states
          and real, imaginary, and complex numbers, all in a
          different reference frame.

          Since there are many different unitary operators $U$ it
          follows that there are many different reference frames
          each with their sets $\mathcal{R}_{U},\mathcal{I}_{U},
          \mathcal{C}_{U}$ of real, imaginary, and complex numbers
          with appropriate definitions of $=_{RU},\leq_{RU},+_{RU},$ etc.

           \section{Gauge Transformations of
           Representations}\label{GTR}
          Gauge transformations of the qubit strings are an interesting
          type of unitary transformation to consider.
          Both global and local gauge transformations of qubit strings
          will be considered. These are
         described by elements of $SU(2)$ acting on each $a$ and $b$
         qubit. Their possible action on the sign $c$ and $d$ qubits
         will be suppressed here to keep things simple.
         Global gauge transformations, which have the same $U$ applied
         to each qubit in a string, have been recently considered
         in the context of reference frame change in quantum
         information theory \cite{Enk} and are used in studying
         decoherence free subspaces \cite{Lidar}.

         Gauge transformations are represented here by
         \begin{equation}\label{Uglsdkt}
         \gsdt\rightarrow U\gsdt=(\times_{j=l_{r}}^{u_{r}}
         U_{j,1}|\g s\rangle)\times(\times_{j=l_{i}}^{u_{i}}
         U_{j,2}|\d t\rangle)\end{equation}
         where\begin{equation}\label{prodU}\begin{array}{l}
         (\times_{j=l_{r}}^{u_{r}} U_{j,1}|\g s\rangle)
         \times(\times_{j=l_{i}}^{u_{i}} U_{j,2}|\d t\rangle) \\
         \hspace{1cm}=\cd_{\g,0}(\ad_{U_{l_{r},1}})_{s(l_{r}),l_{r}}\cdots
         (\ad_{U_{u_{r},1}})_{s(u_{r}),u_{r}} \\ \hspace{1cm}\times\dd_{\d,0}
         (\bd_{U_{l_{i},2}})_{t(l_{i}),l_{i}}\cdots
         (\bd_{U_{u_{i},2}})_{t(u_{i}),u_{i}}|0\rangle.\end{array}
         \end{equation} For global transformations $U_{j,i}$ is the same
         operator for all index values. Transformations of the
         single qubit $a$ and $b$ operators are given by
         \begin{equation}\label{adaU}\begin{array}{c}
         (\ad_{U_{j,1}})_{h,j}=U_{j,1}\ad_{h,j}
         \\(a_{U_{j,1}})_{h,j}=a_{h,j}U_{j,1}^{\dag}.\end{array}
         \end{equation}From the expansion of $U_{j,1}$ as
         \begin{equation}\label{Uexpand}
         U_{j,1}=\sum_{i,h}(U_{j,1})_{i,h}\ad_{i,j}a_{h,j}
         \end{equation} one obtains
         \begin{equation}\label{aUmat}\begin{array}{c}
         (\ad_{U_{j,1}})_{h,j}=\sum_{i=0,1}(U_{j,1})_{i,h}\ad_{i,j}
         \\(a_{U_{j,1}})_{h,j}=
         \sum_{i}(U_{j,1})^{*}_{i,h}a_{i,j}.\end{array} \end{equation} The same
         equations hold with $b$ replacing $a$ and $U_{j,2}$
         replacing $U_{j,1}$ for the
         operators of the imaginary component.

         These results can be inserted into the transformed string
         state $U\gsdt$ to express it in terms of the original
         $\ad_{i,j},\bd_{i^{\p},j^{\p}}$.  The result is a product of terms with
         each term given by Eq. \ref{aUmat}. For example Eq. \ref{aUmat}
         gives $(\ad_{U_{j,1}})_{s(j),j}=(U_{j,1})_{1,s(j)}
         \ad_{1,j}+(U_{j,1})_{0,s(j)}\ad_{0,j}$ and
         $(\bd_{U_{j,2}})_{t(j),j}=(U_{j,2})_{1,t(j)}
         \bd_{1,j}+(U_{j,2})_{0,t(j)}\bd_{0,j}.$

         The arithmetic operators
         $\tilde{+}_{A},\tilde{\times}_{A},\tilde{-}_{A},\tilde{\div}_{A,l}$
         transform under $U$ in the expected way. For $\tilde{+}_{A}$
         one defines $\tilde{+}_{A,U}$ by \begin{equation}\label{addAU}
         \tilde{+}_{A,U}=(U\times U)\tilde{+}_{A}(U^{\dag}\times U^{\dag}).
         \end{equation} Then
         \begin{equation}\label{addAUA}\begin{array}{l}
         \tilde{+}_{A,U}(U\gsdt\times U\gsdtp)\\ \hspace{1cm}
         =(U\times U) \tilde{+}_{A}(\gsdt\times \gsdtp)\end{array}\end{equation}
         as expected. This is consistent with the definition of
         $\tilde{+}_{A}$ in Eq. \ref{addn} which shows that
         $\tilde{+}_{A}$ is a binary relation. Similar relations
         hold for the other three operators.

         The same transformations apply also to the
         arithmetic relations $=_{A}$ and $\leq_{A}.$ One defines
         $=_{A,U}$ and $\leq_{A,U}$ by\begin{equation}\label{=AU}
         \begin{array}{c}=_{A,U}:= U=_{A}U^{\dag} \\
         \leq_{A,U}:= U\leq_{A} U^{\dag}.
         \end{array}\end{equation} These relations express the fact
         that $U\gsdt =_{AU}U\gsdtp$ if and only if
         $\gsdt =_{A}\gsdtp$ and $U\gsdt \leq_{AU}U\gsdtp$ if and only if
         $\gsdt\leq_{A}\gsdtp.$ In terms of projection operators for
         these relations one has\begin{equation}\label{prpjeq}\langle
         U(\g^{\p}s^{\p}\d^{\p}t^{\p})|P_{=_{AU}}U|\g s,\d t\rangle = \langle
         \g^{\p}s^{\p}\d^{\p}t^{\p}|P_{=_{A}}|\g s,\d t\rangle
         \end{equation} where $P_{=_{AU}}=UP_{=_{A}}U^{\dag}.$
         Similar relations hold for $\leq_{AU}$ and $\leq_{A}.$

         The above shows the importance of the subscripts on the
         arithmetic operations and relations.
         $|\cd_{\g,0}\ad_{1,0}\rangle
         =_{A}|\cd_{\g,0}\ad_{1,0}\rangle$ trivially, but
         $U\cd_{\g,0}\ad_{1,0}|0\rangle =_{A}
         U\cd_{\g,0}\ad_{1,0}|0\rangle$ is false even though
         $U\cd_{\g,0}\ad_{1,0}|0\rangle =_{A,U}
         U\cd_{\g,0}\ad_{1,0}|0\rangle.$ This
         is nothing more than the statement that two spin
         systems, which are in state $|\uparrow\uparrow\rangle$
         in one reference frame, have nonzero components
         $|\uparrow\downarrow\rangle$ and $|\downarrow\uparrow\rangle$
         in a rotated reference frame.

         It follows from this that if  $\gsdt$ is a
         rational string state in a reference frame with the
         $A$ arithmetic operations and relations, then
         $U\gsdt$ is a rational string state in the transformed
         frame with the $UA$ arithmetic operations and relations.
         However $\gsdt$ and $U\gsdt$ are not rational string states
         in the transformed and original frames respectively. The reason
         is that the states viewed in the different frames are
         superpositions of many different qubit string states including
         those with leading and trailing $0s,$ which are excluded in the
         representation used here.

         These considerations extend to the real and complex number
         operators. If $\tilde{O}$ is a normalized Cauchy operator
         in one frame, then the transformed operator\begin{equation}\label{defOU}
         \tilde{O}_{U}= U\tilde{O}U^{\dag} \end{equation}
         is Cauchy in the transformed frame.
         However $\tilde{O}_{U}$ is not a Cauchy operator in the original
         frame and $\tilde{O}$ is not Cauchy in the transformed frame.
         To show that $\tilde{O}_{U}$ is Cauchy in the
         transformed frame if and only if $\tilde{O}$ is Cauchy in
         the original frame one can start with the expression for
         the Cauchy condition in the transformed frame:
         \begin{equation}\label{cauchtOU}
         \begin{array}{c}|\tilde{O}_{U}U|s_{j}\rangle
         -_{AU}\tilde{O}_{U}U|s_{k}\rangle|_{AU}\leq_{AU}U|+,-\ell\rangle \\
         \mbox{for all $U|s_{j}\rangle,$ $U|s_{k}\rangle \geq_{AU}$ some
         $U|s_{h}\rangle$}\end{array}\end{equation} From Eq. \ref{defOU}
         one gets $$\tilde{O}_{U}U|j\rangle
         -_{AU}\tilde{O}_{U}U|s_{k}k\rangle=U\tilde{O}|s_{j}\rangle
         -_{AU}U\tilde{O}|s_{k}\rangle.$$ From Eqs. \ref{addn} and
         \ref{resaddn} applied to $-_{AU}$ and Eqs. \ref{addAU} and
         \ref{addAUA} one obtains $$U\tilde{O}|s_{j}\rangle
         -_{AU}U\tilde{O}|s_{k}\rangle =U(\tilde{O}|s_{j}\rangle
         -_{A}\tilde{O}|s_{k}\rangle).$$  Use of\begin{equation}\label{absUA}
         |-|_{AU}=U|-|_{A}U^{\dag} \end{equation}for the absolute
         value operator gives $$|U(\tilde{O}|s_{j}\rangle
         -_{A}\tilde{O}|s_{k}\rangle)|_{AU}=U(|\tilde{O}|s_{j}\rangle
         -_{A}\tilde{O}|s_{k}\rangle|_{A}).$$ Finally from Eq.
         \ref{=AU} one obtains $$\begin{array}{l}U(|\tilde{O}|s_{j}\rangle
         -_{A}\tilde{O}|s_{k}\rangle|_{A})\leq_{AU}U|+,-\ell\rangle
         \\ \hspace{1cm} \leftrightarrow |\tilde{O}|j\rangle
         -_{A}\tilde{O}|s_{k}\rangle|_{A}\leq_{A}|+,-\ell\rangle
         \end{array}$$ which is the desired result. Thus one sees
         that the Cauchy property is preserved in unitary transformations
         from one reference frame to another.

         To show that $\tilde{O}_{U}$ is not Cauchy in the original frame
         it is instructive to consider a simple example.
         First one works with the Cauchy
         property for sequences of states. Let
         $f:(-\infty,n]\rightarrow \{0,1\}$ be a $0-1$ function from
         the set of all integers $\leq n$ where $f(n)=1.$ Define a
         sequence of states\begin{equation}\label{fseq}
         |f\rangle_{m} =\cd_{+,0}\ad_{f(n),n}\ad_{f(n-1),n-1}\cdots
         \ad_{f(-m),-m}|0\rangle\end{equation}for $m=1,2,\cdots.$
         The sequence is Cauchy as $||f(j)\rangle
         -_{A}|f(k)\rangle|_{A}\leq_{A}|+,-\ell\rangle$ for all
         $j,k>\ell.$ However for any  gauge transformation $U$
         the sequence \begin{equation}\label{fseqU}
         U|f\rangle_{m} =\cd_{+,0}(\ad_{U})_{f(n),n}\cdots
         (\ad_{U})_{f(-m),-m}|0\rangle\end{equation} is not Cauchy
         as  expansion of the $\ad_{U}$ in terms of the $\ad$ by Eq.
         \ref{aUmat} gives $U|f\rangle_{m}$ as a sum of terms whose
         arithmetic divergence is independent of $m$.

         For the operator $\tilde{O}$ one has the additional problem
         that applying $\tilde{O}_{U}$ to any integer state
         $|s_{j}\rangle$ is equivalent to applying $U\tilde{O}$ to the
         state $U^{\dag}|s_{j}\rangle.$ However this state is a linear
         sum over many integer states.  This also shows that
         $\tilde{O}_{U}$ is not Cauchy in the original reference
         frame.

         These results can be obtained directly by noting that
         $\tilde{O}_{U}$ can be expressed as a sum of products of
         $a_{U}$ and $b_{U}$ operators and their adjoints. Applying
         this to an original frame state, $|s_{j}\rangle$, which
         is a product of $\ad$ and $\bd$ operators, gives a
         complex linear superposition of states that is not Cauchy.

         It is useful to summarize the results obtained so
         far. For each gauge transformation $U$ there are
         associated sets $\mathcal{R}_{U}, \mathcal{I}_{U},
         \mathcal{C}_{U}$ of Cauchy operators that are real, imaginary,
         and complex numbers in reference frames, one for each $U$
         \cite{Enk}.  All frames are unitarily equivalent in that there
         are unitary transformations that relate the operators in
         the $U$ frame to those in the $U^{\p}$ frame. This includes
         the relations between the operators in
         $\mathcal{R}_{U},\mathcal{I}_{U},\mathcal{C}_{U}$ and in
         $\mathcal{R}_{U^{\p}},\mathcal{I}_{U^{\p}},\mathcal{C}_{U}.$
         Also an operator $\tilde{O}_{U}$ that is Cauchy in the $U$
         frame is not Cauchy in the original frame or in any other $U^{\p}$
         frame.

          \section{Fields of Iterated Quantum Frames} \label{FIQF}

          The organization of the $\mathcal{R}_{U}, \mathcal{I}_{U},
         \mathcal{C}_{U}$ based quantum frames into a field structure
         is based on noting that the whole discussion so far  can be
         considered as taking place in one frame $F$ that is based on
         one set each of real numbers $R$ and complex numbers $C$ and
         one set of continuum $4$ tuples  $R^{4}$  as a space time
         basis. These are considered as external and given in some
         absolute mathematical sense. This frame can also be
         considered to be the basis of reference frames described in
         the literature, such as those in \cite{Aharonov}-\cite{vanEnk},
         \cite{Enk}.

         It is clear from the above that $F$ is sufficiently
         extensive to serve as a basis for all physical theories as
         mathematical structures based on $R$ and $C$.  This includes special
         and general relativity, quantum mechanics and quantum
         field theory, QED and QCD, and string theory (if $R^{4}$ is
         replaced by $R^{D}$ where $D>4.)$ Loop quantum gravity is
         included  to the extent that it is based on the real and
         complex numbers. Similar arguments apply to space and spin
         foam approaches \cite{Ng,Girelli,Terno,Perez} to this
         topic.

         The previous discussion has also shown that $F$ includes a
         quantum theory description of complex rational string states
         of qubit strings.  It  includes arithmetic relations as
         properties of these states and arithmetic operators on
         these states. It also contains Cauchy operators, $\tilde{O},$
         which satisfy the relevant properties of real, imaginary, and
         complex numbers.  In addition $F$ includes all local and global gauge
         transformations $U$ of the complex rational string states and
         the corresponding Cauchy operators $\tilde{O}_{U}.$ Also
         included is a description of the physical dynamics of the
         string of systems whose states are the complex rational
         string states.

         An important consequence of these observations is that the
         frame $F$ also includes many other frames. For each
         gauge $U$ one has a frame $F_{U}$ that is based on the real
         and complex numbers $\mathcal{R}_{U},\mathcal{C}_{U}$ and a
         space time $\mathcal{R}_{U}^{4}$  that are sets of the Cauchy
         operators $\tilde{O}_{U}.$ Corresponding to the basic
         relations $=,\leq$ and operations $+,\times ,-,\div$  and derived
         relations of mathematical analysis on $R$ and $C,$  in
         frame $F$ are the relations $=_{RU},\leq_{RU}$ and operations
         $+_{RU},\times_{RU} ,-_{RU},\div_{RU}$  and derived
         relations of mathematical analysis on $\mathcal{R}_{U}$ and
         $\mathcal{C}_{U},$  in frame $F_{U}.$ Similar
         correspondences follow in that all properties of space time
         $R^{4}$ in $F$ should have corresponding properties of
         space time $\tilde{R}_{U}^{4}$ in $F_{U}.$

         It follows that $F_{U}$
         is equivalent to $F$ in that much, and possibly almost all, of the
         physics and math described in $F$ by an observer in $F$ can
         be mapped isomorphically into $F_{U}.$ The ultimate
         basis for this is the existence of  isomorphisms between
         $R$ and $\mathcal{R}_{U},$  $C$ and $\mathcal{C}_{U},$ and
         $R^{4}$ and $\mathcal{R}^{4}_{U},$ and the observation that
         physical theories are mathematical structures based on the
         real and complex numbers.

         As an example of the equivalence,
         let \begin{equation}\label{physlaw} G(X_{1}\cdots
         X_{n})=0\end{equation} represent a physical law in frame
         $F$. Here $G$ is some function of the $n$ variables and
         constants, denoted collectively by $X_{1},\cdots X_{n},$
         that take values in $R,C,$ and $R^{4}$ and in other
         physical variables. The structure of $G$ is ultimately
         based on the basic and derived relations for $R$, $C$ and
         $R^{4}.$ This law has a corresponding expression in $F_{U}:$
          \begin{equation}\label{physlawU}
         G_{U}((X_{U})_{1}\cdots (X_{U})_{n})=_{RU}0_{U}.
         \end{equation} Here $G_{U}$ is obtained from $G$ by
         replacements $=\rightarrow =_{RU},\leq\rightarrow \leq_{RU},$
         $+\rightarrow +_{RU},\times\rightarrow \times_{RU},$ and
         $-\rightarrow -_{RU},\div\rightarrow \div_{RU}.$ Also
         $0_{U}$ is a Cauchy operator that is the number $0$ in
         $\mathcal{R}_{U}$. This also assumes that the ranges of the
         variables and values of the constants appearing in Eq.
         \ref{physlaw} have correspondents in $F_{U}$.

         In addition all frames $F_{U}$ are unitarily equivalent
         regarding their descriptions of physics and math.  This can
         be seen by noting that in the frame $F$ the sets
         $\mathcal{R}_{U}$ and $\mathcal{R}_{U^{\p}},$ and
         $\mathcal{C}_{U}$ and $\mathcal{C}_{U^{\p}},$  where
         $U$ and $U^{\p}$ are two gauge transformations, are related by
         a unitary transformation.  In particular one has
         \begin{equation}\label{OUUpp} O_{U^{\p}}=U^{\p\p}O_{U}
         (U^{\p\p})^{\dag} \end{equation} where
         $U^{\p\p}=U^{\p}U^{\dag}.$

         Note also that the frame $F_{ID}$ where $U$ is the identity
         is one of the frames $F_{U}$.  It is \emph{not} the same as
         the frame $F$ as $\mathcal{R}_{ID}=\mathcal{R}$ and
         $\mathcal{C}_{ID}=\mathcal{C}$ are (equivalence classes of)
         the Cauchy operators $\tilde{O}.$ It is also the case that
         the choice of which gauge transformation $U$ is the identity
         depends on a corresponding choice of a rational string state
         basis in $F$. Since this is arbitrary and all bases are equivalent,
         the choice is much like gauge fixing in quantum field theory.

\begin{figure}[h]\begin{center}
           \resizebox{120pt}{120pt}{\includegraphics[230pt,200pt]
           [490pt,460pt]{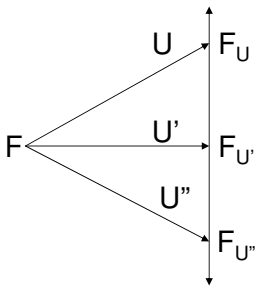}}\end{center}
           \caption{Relation between a Base Frame $F$ and Gauge
           Transformation Frames. A frame $F_{U}$ is associated
           with each gauge transformation $U$ of the rational string
           states in $F$. The three frame connections shown are
           illustrative of the infinitely many connections,
           shown by the two headed vertical arrow. Each $F_{U}$
           is based on real and complex numbers $\mathcal{R}_{U},
           \mathcal{C}_{U}$ and a space time arena $\mathcal{R}^{4}_{U}$.}
           \label{RCST1} \end{figure}

         The relationship between frame  $F$ and the frames
         $F_{U}$ is shown in Fig \ref{RCST1}. All $F_{U}$ with
         $\mathcal{R}_{U},\mathcal{C}_{U}$ and space time
         $\mathcal{R}^{4}_{U}$ are seen emanating
         from the base or parent frame $F$. The three
         frames shown with arrows are examples of the connections from
         $F$ to each of the infinitely many
         frames, exemplified by the solid vertical double arrow
         as there is one frame associated with each $U$.

         There is an asymmetric relation between the parent frame
         $F$ and the $F_{U}$ frames.  An observer in $F$ can view
         the $F_{U}$ frames and see and describe their construction.  He
         can see that for each $U,$ $\mathcal{R}_{U}$ and $\mathcal{C}_{U},$
         as sets of Cauchy operators on rational string states, and
         $\mathcal{R}_{U}^{4}$ are the real and complex numbers and space
         time arena that are the basis of $F_{U}$.

         An observer in $F_{U}$ has a different view.  To him
         $\mathcal{R}_{U},\mathcal{C}_{U}$ are abstract sets just as
         $R$ and $C$ are abstract to an observer in $F$.  The observer
         in $F_{U}$ cannot see  $\mathcal{R}_{U}$ and $\mathcal{C}_{U}$
         as Cauchy operators in an underlying frame. Similarly he views
         space time, $\mathcal{R}^{4}_{U},$ as a $4-$tuple of abstract real
         numbers.  This limit on his vision occurs because
         the observation that $R_{U},C_{U}$  are sets of Cauchy
         operators is external to or outside of his frame. In this way
         his view of the real and complex numbers and space time is
         the same as that of an observer in $F$ about his
         own real and complex numbers and space time.  In addition,
         the observer in $F_{U}$ cannot see the frame $F.$ This is a
         result of the abstract, external nature of $\mathcal{R}_{U},
         \mathcal{C}_{U},$ and $\mathcal{R}^{4}_{U}.$

         Possibly the most important aspect of the frame relations
         shown in Fig. \ref{RCST1} is that the construction can be
         iterated.  This is based on noting that the definitions of
         binary complex rational string states,  the basic arithmetic
         relations $=_{A}$ and $\leq_{A}$ and operations
         $+_{A},\times_{A},-_{A},\div_{\ell,A}$, Cauchy conditions,
         and definitions of the Cauchy operators, all have their
         equivalent definitions in $F_{U}.$ Note that a rational
         string state $|\g s,\d t\rangle_{U}$ in $F_{U}$ is not the
         same as either the state  $|\g s,\d,t\rangle$ or
         $U|\g s,\d,t\rangle.$

\begin{figure}[h]\begin{center}
           \resizebox{130pt}{130pt}{\includegraphics[250pt,160pt]
           [540pt,450pt]{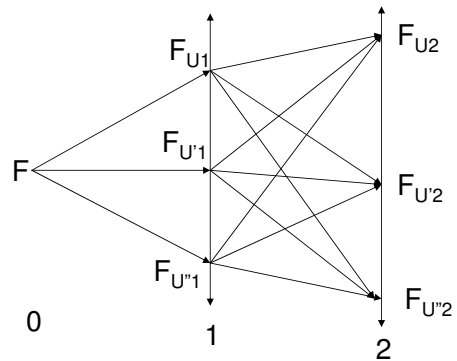}}\end{center}
           \caption{Three Iteration Stages of Frames coming from
           Frames. Only three frames of the infinitely many, one for
           each gauge $U,$ are shown at stages $1$ and $2$. The
           arrows connecting the frames show the iteration direction
           of frames emanating from frames.} \label{RCST2} \end{figure}

         This iteration is illustrated in Fig. \ref{RCST2} that
         shows two stages of the iteration as well as the base frame
         $F$ at stage $0$.  At each stage only three of the infinite
         number of frames, one for each gauge $U$, coming from each
         parent are shown.  One sees that each frame in stage $2$ has
         an infinite number of parents, one for each gauge U, and
         each frame in stage $1$ has an infinite number of children,
         one for each gauge $U$.

\begin{figure}[h]\begin{center}
           \resizebox{130pt}{130pt}{\includegraphics[230pt,120pt]
           [560pt,490pt]{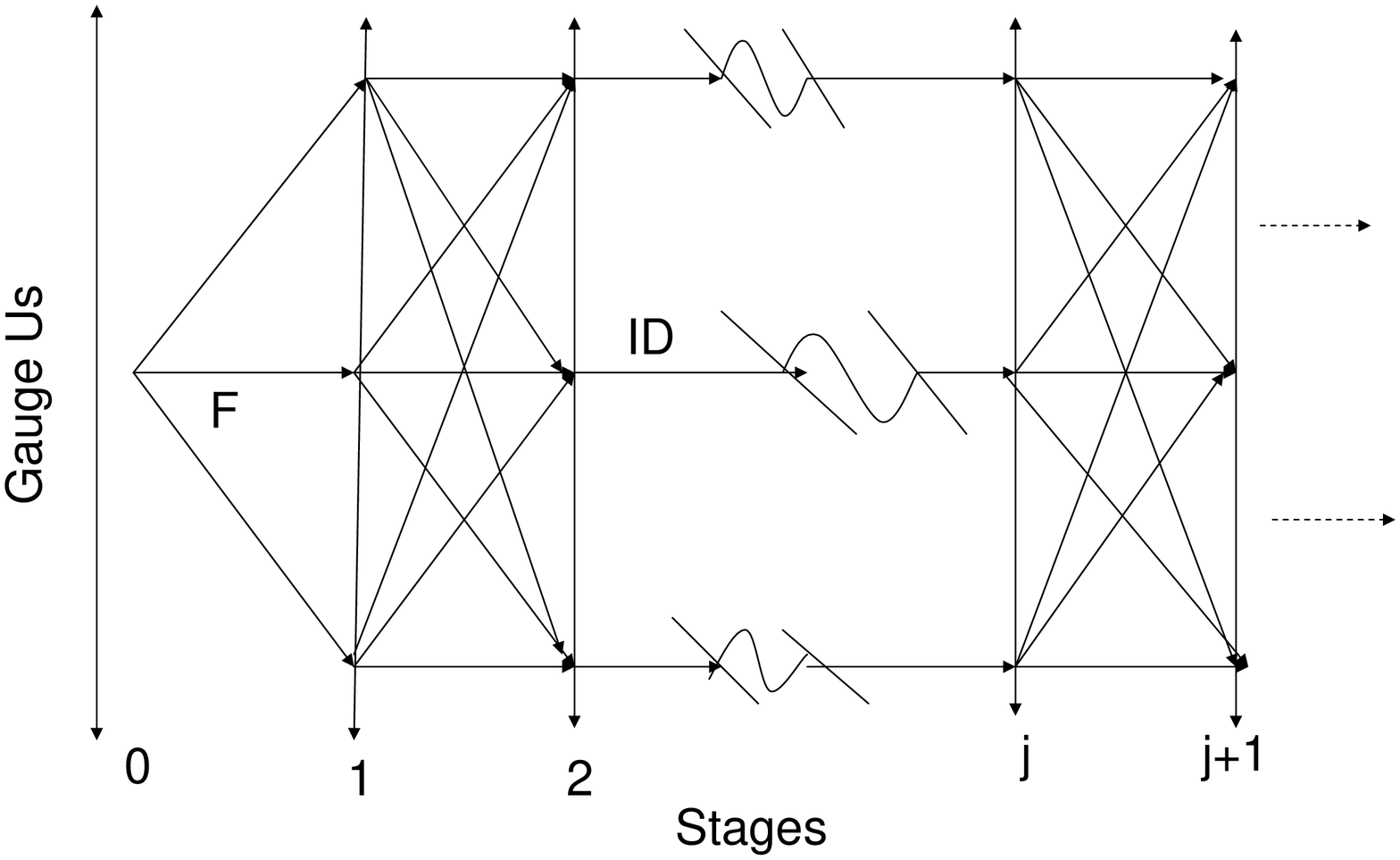}}\end{center}
           \caption{One way Infinite Iteration of Frames coming from
           Frames. Only three of the infinitely many frames, one for
           each gauge $U$ are shown for stages $1,2,\cdots,j,j+1,\cdots.$
           The arrows connecting the frames show the iteration or emanation
           direction. The center arrows labeled ID denote iteration of the
           identity gauge transformation.} \label{RCST3}
           \end{figure}

         It is also clear from the preceding that this iteration can
         be continued indefinitely to give stages $0,1,2,\cdots.$
         Each stage $j$ contains an infinite number of frames
         $(F_{j})_{U}$, one for each gauge $U$. Each $(F_{j})_{U}$ has
         an infinite number of children, $(F_{j+1})_{U},$ one for
         each $U$, and an infinite number of parents,
         $(F_{j-1})_{U}$, also one for each gauge $U$. Also  any
         frame  $(F_{j})_{U}$ at any level $j$ can serve as the
         origin of a one way infinite directed structure of frames
         emanating from frames just as the frame $F$ does  in Fig.
         \ref{RCST3}.
           This figure also shows that iterated frame structure can
           be viewed as a field of frames on a two dimensional
           background of discrete stages in one direction and
           continuous gauge transformations in the other direction.
           Each point on each of the two headed arrows corresponds
           to a frame. The abcissa and ordinate denote respectively
           the discrete stages and the gauge transformations. The
           central arrows labeled "ID" denote the identity gauge at each
           stage.  Its location on the gauge axis depends on the arbitrary
           choice made in $F$ to represent the complex binary rational
           string states. Since an observer in $F$ can see all frames
           downstream  he can determine which gauge at each stage is
           the identity.

           One also notes that the whole structure shown in Fig.
           \ref{RCST3} can be translated to any frame.  That is, any
           frame at any stage, can serve as the origin of a
           coordinate system just as $F$ does in the figure. In this
           case the arbitrary choice in the new frame of the gauge
           for the complex rational string states is the identity
           gauge for the new set up.  In some ways this is similar
           to coordinate frame translations and rotations in usual
           3D space.

           The frame field in  Fig. \ref{RCST3} has the problem
           that not all frames are equivalent. The base frame $F$
           is unique or special in that for it $R$, $C$, and space
           time $R^{4}$, are external, or abstract and given. They
           are not sets of operators in any frame. These aspects of
           $F$ are outside the whole frame field of the figure and
           would have to be meaningful in some absolute sense,
           whatever that is.

\begin{figure}[h]\begin{center}
           \resizebox{130pt}{130pt}{\includegraphics[230pt,120pt]
           [560pt,490pt]{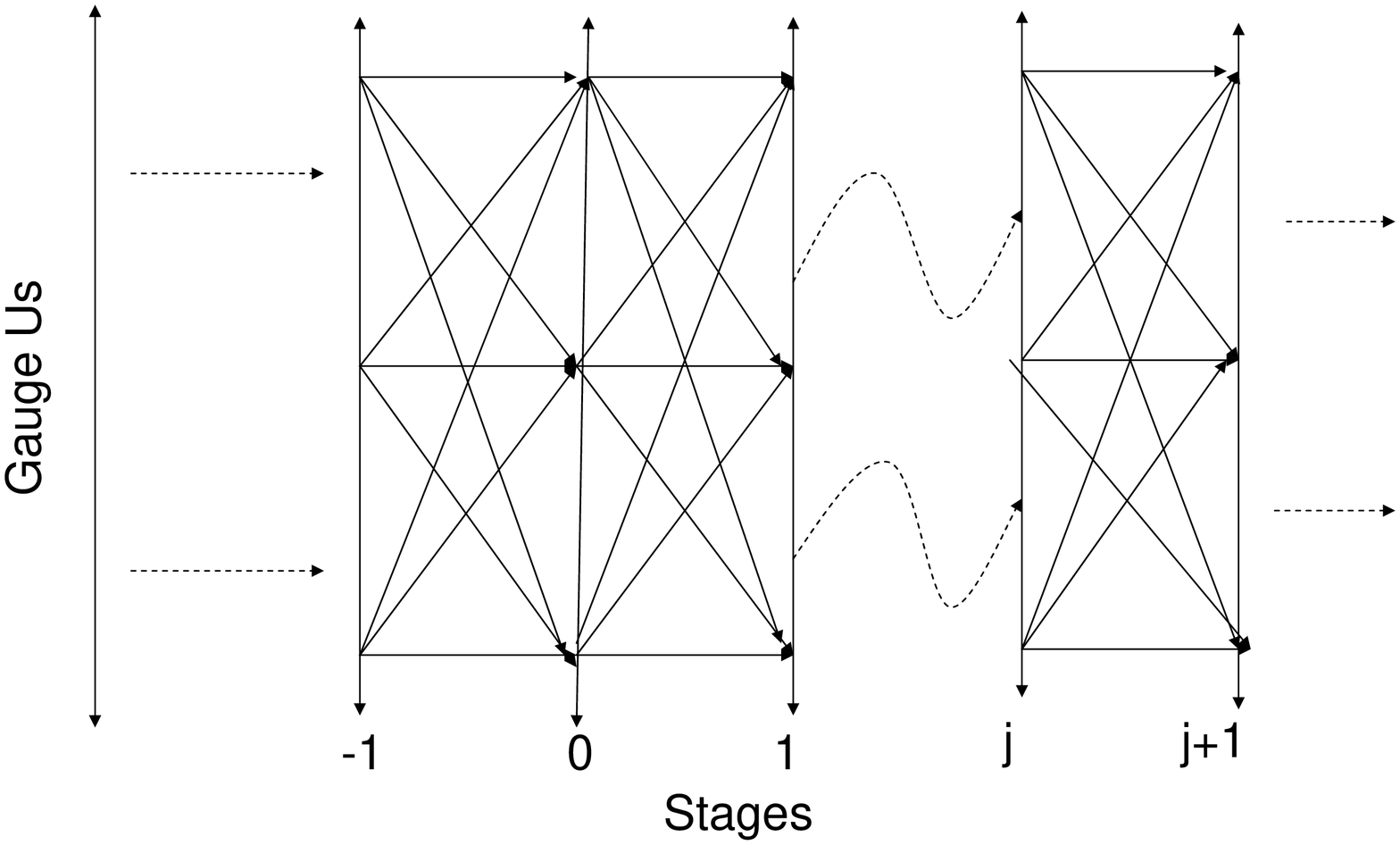}}\end{center}
           \caption{Two way Infinite Iteration of Frames coming from
           Frames. Only three of the infinitely many frames, one for
           each gauge $U$ are shown at each stage $\cdots ,-1,0,1,
           \cdots,j,j+1,\cdots$. The solid arrows connecting the
           frames show the iteration or emanation direction. The wavy
           arrows denote iterations connecting stage $1$ frames to
           those at stage $j$.  The straight dashed arrows denote
           infinite iterations from the left and to the right.of the
           identity gauge transformation.} \label{RCST4} \end{figure}

           One way to escape this problem is to extend the frame
           field to infinity in both stage directions. This is shown
           in Fig. \ref{RCST4} where stages are labeled with all
           integers. In this case there is no special frame as each
           frame has an infinite number of children and parents,
           one child and one parent for each gauge $U$. The frame
           field retains the basic direction of frames emanating from
           frames as expressed by the parent child relationship.

           This provides a satisfactory solution to the problem of
           the meaning of the existence of $R,C$, and $R^{4}$ in
           some absolute sense.  For the two way infinite frame
           field there are no external or abstract $R,C,$ and
           $R^{4}.$ The frame field is background independent in
           that there is no space time or real or complex number
           fields that exist in an absolute sense.

           Instead these exist in a relative sense only in that the
           real and complex number basis for each frame in the field
           are external and abstract for an observer in the frame.
           They are not external or abstract for an observer in any
           parent frame as they are consist of sets of Cauchy
           operators based on gauge transformations of
           complex rational string states in the parent frame. In
           this way the notion of idealized mathematical existence
           of real and complex numbers and of space time is relative
           as it is internal to the frame field. It has meaning only
           for observers within frames. This point will be taken up
           again in Section \ref{CP} on the connection of the frame
           field to physics.

           \section{String Number Bases}\label{SNB}
           \subsection{$k\geq 2$}
           So far, the treatment has been based on qubit strings.
           These correspond to binary representation of numbers.
           However, from a mathematical or physical point of view,
           there is nothing special about
           the base $2$.  Any base $k$ with $k\geq 2$ is just as
           good as any other. Use of the base $2$ as the
           basis of computations may have computational advantages, but
           these do not seem to be relevant to physics in any basic sense.

           It follows that the restriction to $k=2$ should be
           removed.  The approach taken here to accomplish
           this is based on what might be called the maximum
           efficiency of representing all rational numbers as the
           set of all finite digit strings in a base $k$. Note that
           the representation of a real or imaginary rational number
           is by a \emph{single} string state, not by pairs of string
           states.

           The problem at hand is to determine the smallest base, $k_{min}$
           for representing the rational number $1/n$ exactly as a
           finite base $k$ digit string.  It is easy to convince
           oneself that if $n$ is a prime number, then $k_{min}=n.$  If
           $n=p_{1}^{h_{1}}p_{2}^{h_{2}}\cdots p_{m}^{h_{m}}$ is a
           product of $m$ primes to powers $h_{1}\cdots h_{m}$, then
           $k_{min}=p_{1}p_{2}\cdots p_{m}$ is the product of the $m$
           primes all to the first power. It is sufficient to
           consider representations of $1/n$ because any base $k$ that
           can represent $1/n$ exactly can also represent
           $j/n$ exactly where $j$ is any integer. For $n=2,3,\cdots$
           the corresponding pairs $(n,k_{min})$ are
           $(2,2),(3,3),(4,2),(5,5),(6,6)(7,7),(8,2),(9,3),
           (10,10)$, etc. One sees from this that the sequence of
           increasing minimal bases is given by $k_{min}=p_{1}\cdots
           p_{m}$ the product of the first $m$ primes for
           $m=2,3,5,\cdots.$

           These aspects can be incorporated into the use of complex
           rational string states by expanding the qubit base to a
           qukit base where\begin{equation}\label{kprimepd}
           k=p_{1}p_{2}\cdots p_{m}.\end{equation} There are two
           options.  One is to consider qukits as single systems
           with $k$ internal states.  These would be represented
           here by AC operators $\ad_{\a,j},a_{\a,j},\bd_{\b,j},
           b_{\b,j}$ where $\a,\b$ range over $k$ system states.

           The other option, which will be used here, is to consider a
           base $k$ qukit where $k$ satisfies Eq. \ref{kprimepd} as
           an $m$ tuple of qupits , $qu2it\times qu3it\cdots\times
           qu(p_{m})it,$ as a product of $m$  systems with the $j$
           system type having $p_{j}$ internal states. These would
           be represented by $m$ different $a$ and $b$  operators
           $(\ad_{1})_{\a_{1},j},(\ad_{m})_{\a_{m},j},$
           $(\bd_{1})_{\b_{1},j},\cdots ,(\bd_{m})_{\b_{m},j}$ and
           similarly for the annihilation operators. Here
           $\a_{j},\b_{j}$ refer to $p_{j}$ different states of the
           type $j$ system.

           In this case the individual creation operators appearing in
           Eqs. \ref{binranum} and \ref{shorthand} that define the
           complex rational string states $\gsdt$ are replaced by a
           product of the $m$ different types.  That is,
           $\ad_{s(j),j}$  and $\bd_{t(j),j}$ are replaced by
           $\prod_{h=1}^{m}(\ad_{h})_{s_{h}(j),j}$ and
           $\prod_{h=1}^{m}(\bd_{h})_{t_{h}(j),j}$ in these
           equations. In this case a state $\gsdt$ is given by
           \begin{equation}\label{gsdtmprod}
           \gsdt=\cd_{\g,0}\prod_{j=l_{r}}^{u_{r}}\prod_{h=1}^{m}
           (\ad_{h})_{s_{h}(j),j}\dd_{\d,0}\prod_{j=l_{i}}^{u_{i}}
           \prod_{h=1}^{m}(\bd_{h})_{t_{h}(j),j}|0\rangle.
           \end{equation} Here $s_{h}$ and $t_{h}$ are functions
           from their respective domains $[l_{r},u_{r}],[l_{i},u_{i}]$
           to $\{0,1,\cdots ,p_{h}-1\}.$ Alternatively this state
           could be represented as the  product of the states of m
           $qup_{j}it$ strings, one for each $j=1,2,\cdots,m:$
           \begin{equation}\label{gsdtmprod2}
           \gsdt=\times_{h=1}^{m}[\cd_{\g,0}\prod_{j=l_{r}}^{u_{r}}
           (\ad_{h})_{s_{h}(j),j}\dd_{\d,0}\prod_{j=l_{i}}^{u_{i}}
           (\bd_{h})_{t_{h}(j),j}|0\rangle ].
           \end{equation}

           Definitions of the arithmetic relations, $=_{A}$ and
           $\leq_{A},$ and operations $+_{A},\times_{A},$ $-_{A},
           \div_{A,\ell}$ are more complex but nothing new
           is involved. It is similar to changes needed in replacing
           single digits which can range from $0$ to $k-1$ to an $m$
           tuple of digits where the $hth$ ranges from $0$ to
           $p_{h}-1.$ Similarly the definitions of the operators
           $\tilde{O}$ and their Cauchy convergence can be extended
           to apply here. In this case the nonnegative integer
           states  $|\g s\rangle$ become $|\g s\rangle =\cd_{+,0}
           \prod_{h=1}^{m}(\ad_{h})^{s_{h}}_{[0,u_{r}]}|0\rangle$
           with the Cauchy condition given by Eq. \ref{cauchyO}.

           As was the case for $k=2$ these extended Cauchy operators
           can be collected into equivalence classes that are real
           and complex numbers $ \mathcal{R}_{k},\mathcal{C}_{k}$. Also
           $\mathcal{R}_{k}^{4}$ can be used to represent space time.
           The resulting reference frames for different values of $k$
           are all isomorphic to one another as $\mathcal{R}_{k}$
           and $\mathcal{C}_{k}$  are isomorphic to
           $\mathcal{R}_{k^{\p}}$ and $\mathcal{C}_{k^{\p}}.$

           Local and global gauge transformations are extended here
           from being elements of $SU(2)$ to elements of the direct
           product group $U(1)\times SU(2)\times SU(3)\times SU(5)\cdots\times
           SU(p_{m}).$ For each element $U=U_{1}\times U_{2}\cdots
           \times U_{m}$ of this group one can describe corresponding
           real and complex numbers and space time $R_{U},C_{U},$ and
           $R^{4}_{U}.$ The reason $U(1)$ is present will be discussed
           shortly. It is interesting to speculate that if the
           prime numbers correspond to the number of
           projections of spin systems then the group product above
           corresponds to spins $0,1/2,1,2,3,5,\cdots, (p_{m}-1)/2.$
           All these systems are bosons except the spin $1/2$
           systems which are fermions. Also the local gauge group
           $U(1)\times SU(2)\times SU(3)$ plays an important role in
           the standard model of physics \cite{Novaes,Cottingham}.

           \subsection{$k=1$}
           Unary representations of numbers are not usually
           considered because the length of strings of one symbol needed to
           represent a number $n$ is proportional to $n$ instead of
           $\log{n}.$ Also all arithmetic operations are
           exponentially hard. However this representation should
           be included here because $1$ is a prime number and
           because it is present in an essential way.

           It is of interest to note that base restrictions for
           expressing rational numbers, $1/n,$ as finite strings in
           different bases does not apply to integers. All integers
           can be expressed as finite strings in any base.  It is
           tempting to ascribe this to the observations that $1$ is a
           prime factor of any integer and that integers seem to be
           the only type of number expressible in an unary
           representation.

           The reason the unary representation is present in an
           essential way is that it is the only number
           representation that is extensive. Any finite collection
           of physical systems is itself an unary representation of
           a number, the number of objects in the collection.  Also
           the magnitude of any  extensive physical property of a
           system which changes in discrete steps is an unary
           representation of a number, the number of steps in the magnitude.

           In the same sense a string of qubits is an unary
           representation of a number that is the number of qubits
           in the string. This is the case whatever states the qubit
           string is in, including complex rational string states.
           It follows that one should include global and local
           $U(1)$ gauge transformations. These correspond
           to transformations on the AC operators in Eq. \ref{binranum}
           given by \begin{equation}\label{U1AC}
           \begin{array}{c}\ad_{\a,j}\rightarrow U_{j}\ad_{\a,j}
           =e^{i\theta_{j}}\ad_{\a,j} \\ \bd_{\b,k}\rightarrow
           U_{k}\bd_{\b,k} = e^{i\phi_{k}\bd_{\b,k}}.\end{array}
           \end{equation} If $\theta_{j}$ and $\phi_{k}$  are
           independent of $j$ and $k$ then $U$ is a global gauge
           transformation.  Otherwise it is local.

           \section{Connection to Physics}\label{CP}
           \subsection{General Aspects}
           It is useful at this point to define a  frame stage
           operator $W(U)$, parameterized by $U,$ that takes a
           frame $F_{U^{\p}}$ at stage $j$ to a frame $F_{U^{\p\p}}$
           at stage $j+1:$ \begin{equation}\label{Frtrans}
           F_{U^{\p\p}}=W(U)F_{U^{\p}} \end{equation} where
           $U^{\p\p}=UU^{\p}.$ For any path $F_{U_{j}},F_{U_{j+1}},
           \cdots , F_{U_{j+h}}$ of frames from stage $j$ to stage
           $j+h$ there is an associated product of frame stage
           operators that relates the initial and final frames by
           \begin{equation}\label{framepath} F_{U_{j+h}}=W(U^{\p}_{j+h})
           W(U^{\p}_{j+h-1})\cdots W(U^{\p}_{j+1}F_{U_{j}}.
            \end{equation}  Here \begin{equation}\label{Urel}
            U_{j+k}=U^{\p}_{j+k}U_{j+k-1} \end{equation} for
            $k=1,\cdots, h.$

            It follows from the properties of frames in the frame
            field that $W(U)$ can be expressed as the product of an
            isomorphism $I$ and a gauge dependent operator $V(U)$ as
            $W(U)=V(U)I.$ Here $I =W(ID)$ is the stage step operator
            corresponding to the identity gauge  and $V(U)$
            changes a frame $F_{U^{\p}}$ at any stage to a frame
            $F_{U^{\p\p}}$ at the same stage. Since all frames at any
            stage are unitarily equivalent, one sees that any pair of
            frames in the field are equivalent as far as physics is
            concerned. The physical descriptions and dynamics of
            systems on one frame is related to that in another frame
            by a product of isomorphisms  $W(ID)$ and unitary
            equivalence maps $V(U).$ However this is quite different
            from saying the physics is the same in each frame.

           This is in direct contradiction with our experience.
           Physically there is only one space time as $R^{4},$ and
           there is one set each of abstract real and complex numbers. The
           space time arena in which physics is done includes all of
           space time and it covers cosmological as well as Planck
           aspects.  There is no multitude of space times and of real
           and complex numbers which are unitarily  and isomorphically
           related.

           The consequence of this is that there is an important
           ingredient left out of the treatment so far.  This is
           that one must investigate various methods to make  the
           physics in all frames the same, not just unitarily and
           isomorphically equivalent. If this can be achieved then
           all frames in the frame field describe the same physics
           which, hopefully, is the physics that is observed. Note
           that it is sufficient to show that any one frame can be
           identified with a parent frame. This is sufficient to
           collapse the frame field to one frame.

           Details of how to achieve this must await future work. They may
           include quantum to classical limits by letting the qubits
           become increasingly complex, massive, and more
           classical, taking $m$ on $p_{m},$ the $mth$ prime,
           to infinity, and many other possibilities. One should
           also note that the existence of superselection rules
           corresponds to a reduction of frames as there are no
           frames connecting superselected subspaces of states
           \cite{Enk,AharonovII}.

            Another point concerns experimental support for the frame
           picture shown here. This is especially relevant because
           our experience does not show in any direct way the existence
           of many frames. One possible way this might occur is for
           a physical prediction or theory to arise from the
           frame picture that has not been possible so far.

           A speculative potential candidate is a new approach to the
           unification of gravity with quantum mechanics. This is based
           on the observation that  the frame structure shown here allows
           a different approach to this problem and possibly to other
           problems. So far, attempts to solve this problem, such as
           string theory, correspond here to an observer in a
           frame $F$ trying to unite quantum mechanics with gravity
           for the space time and real, $R$, and complex, $C$,
           numbers of his own frame. One might expect problems with this since
           $R$, $C$, and the space time $R^{4}$ are abstract and
           given entities for $F.$  All properties they may have,  that
           derive from their origin as operators on physical systems,
           are external to $F.$  They are available to an observer in
           a parent or ancestor frame but not to an observer in $F$.

           This results in the possibility that some physical properties
           of Cauchy operators that are based on systems in rational string
           states  in a parent frame could be interpreted by an observer in
           frame $F$ as the existence of gravity for his background space
           time.  This avoids the situation of an observer having to
           tie gravity with quantum mechanics for his own space
           time.  Also since every frame has parent and ancestor
           frames (Fig. \ref{RCST4}) observers in all frames would
           report the existence of gravity in their physical
           universe. At this point, this idea remains to be verified.
           However, if it is successful, it would help to support the
           view presented here.

           It also is worth noting that the two way
           infinite frame field, Fig. \ref{RCST4}, shares
           a feature in common with loop quantum gravity
           \cite{Smolin,Rovelli}.  This is that they are both space time
           background independent. Here he background independence
           is based on the observation that there is no fixed
           external space time background associated with the
           frame field.  Space times are backgrounds only in a relative
           sense. Each frame has a space time that is a background to
           an observer in the frame.  However there are as many space
           time backgrounds as there are frames in the  frame field.
           No one is privileged over another. There is no space time
           background for the field itself.

           \subsection{The Decoherent Free Subspace
           Approach}\label{DFSA}

           One approach to connecting the frame field to the one
           observed frame of physics is based on changing the definition of
           qubits in a way so the qubit states are invariant under a
           group of some gauge transformations. This has the result
           of collapsing all frames  at any stage $j$ that are
           distinguished by any gauge $U$ in the group into one stage
           $j$ frame. If this could be extended to all gauge $U,$ then all
           stage $j$ frames could be collapsed into just one frame at $j$.

            This idea is already in use in a slightly different
            context in  the theory of decoherence subspaces (DFS)
           \cite{Lidar,LidarII}. The goal of the DFS method of
           constructing quantum error correction codes is to exhibit
           two subspaces of states of two or more physical qubits that
           are invariant under different types of errors. States in
           one subspace are the $0$ logical qubit states. States in
           the other are the $1$ logical qubit states.

           This distinction between logical and physical qubits has
           not been made so far in that logical qubit states have
           been identified with states of physical systems as
           physical qubit states. In general, however, logical
           qubit states, correspond to projection operators
           $P_{0_{L}}P_{1_{L}}$ on two subspaces of states of two
           or more physical qubits.  Any pure state of physical
           qubits in the $P_{0_{L}}$ or $P_{1_{L}}$
          subspace corresponds respectively to a logical qubit state
           $|0\rangle_{L}$ or $|1\rangle_{L}$. Mixed states in the
           subspaces would be represented by density operators
           $\rho_{0_{L}},\rho_{1_{L}}.$

           In this more general case the whole construction presented
           here of complex rational string states, real and complex
           numbers as Cauchy operators, and different frames $F_{U}$
           with associated space times $\mathcal{R}^{4}_{U},$ would
           be expanded to describe rational string states in terms of
           strings of the logical qubit subspace projection operators.
           The advantage of this approach is that, logical qubit
           subspaces can be defined that are invariant under the action
           of global gauge $U$ acting on the physical qubit states.

           The DFS approach \cite{Lidar,LidarII} to quantum error
           avoidance in quantum computations makes use of this generalization.
           In this approach quantum errors
           correspond to gauge transformations $U$, so the goal is
           to find subspaces of states that are invariant under at least
           some gauge $U$. One way to achieve this for  qubits is based on
           irreducible representations of direct products of $SU(2)$ as
           the irreducible subspaces are invariant under the action of
           some $U$. Here, to keep things simple, the discussion is
           limited to binary rational string states $(k=2)$.

           As an example, one can show that \cite{Enk}
           the subspaces defined by the irreducible 4 dimensional
           representation of $SU(2)\times SU(2)$ are invariant under
           the action of any global $U$. The subspaces are the three
           dimensional subspace with isospin $I=1$, spanned by the states
           $|00\rangle,|11\rangle,1/\sqrt{2}(|01\rangle +|10\rangle)$
           and the $I=0$ subspace containing $1/\sqrt{2}(|01\rangle
           -|10\rangle).$ States in these subspaces are logical
           qubit states. The action of any global $U$ on
           states in the $I=1$ subspace can change one of the
           $I_{z}$ states into linear superpositions of all states in
           the subspace.  But it does not connect the states in the $I=1$
           subspace with that in the $I=0$ subspace.

            It follows that replacement of the physical qubit states
           in the complex rational string state by logical qubit states
           from the invariant subspaces, gives the result that, for
           all global $U$, the $F_{U}$ frames at any stage $j$ all become
           just one frame at stage $j$. The price for this frame reduction
           is the increased entanglement complexity of logical qubit
           states in a parent frame at stage $j-1.$

          This process can be extended in several ways. One way is
          to consider logical qubits whose states are superpositions
          of states of $3$ physical qubits \cite{Enk,Lidar}.  In this
          case the subspaces of interest are the $8$ dimensional
          irreducible subspaces of $SU(2)\times SU(2)\times SU(2)$.
          These correspond to one $I=3/2$ subspace and two $I=1/2$
          subspaces of $3$ physical qubit states. In \cite{Enk,Lidar}
          the two $I=1/2$ subspaces are taken to be the logical
          qubit $0$ and $1$ subspaces.

          \section{Summary and Discussion}\label{SD}

          In this work the basic importance of real and complex
          numbers to physics  is emphasized.  All physical theories
          are mathematical structures over the real and complex
          numbers, and space time is based on a 4 tuple of the real
          numbers. These aspects have been used to define fields of
          iterated quantum frames that are based on underlying
          complex rational string states and their gauge transformations.
          Rational string states are defined as products of AC operators
          acting on the vacuum string state. Sequences of these states are
          replaced by operators $\tilde{O}$ acting on the string states
          corresponding to nonnegative integers.

          The Cauchy condition for sequences of these states is used
          to define Cauchy operators. (Equivalence classes of) These
          operators are real and complex numbers in the same sense
          as elements of $R$ and $C$ which are the basis of the
          Fock space in which the AC operators are defined. The sets
          $\mathcal{R},\mathcal{C}$ and $\mathcal{R}^{4}$ of Cauchy
          operators are the real and complex numbers and space time
          that are the basis of a stage $1$ frame $F_{1}$.

          The frame base is greatly expanded by gauge
          transformations on the rational string states.  For each
          gauge $U$ one has sets of Cauchy operators $\tilde{O}_{U}$
          that correspond to real $\mathcal{R}_{U}$ and complex
          $\mathcal{C}_{U}$ numbers and space time $\mathcal{R}^{4}_{U}.$
          These form the basis for an infinite number of stage $1$
          frames $F_{U}$, one for each $U$ (Fig. \ref{RCST1}). All
          these frames are equivalent as they are related by unitary
          transformations. $F_{1}=F_{ID}$ where $ID$ is the
          identity.

          This process of frame construction can be iterated in one
          or two directions.  This gives either one way infinite
          (Fig. \ref{RCST3}) or two way infinite (Fig. \ref{RCST4})
          fields of frames coming from frames. For a one way infinite field,
          each frame has an infinite number of children frames, one
          for each $U$. For a two way infinite field, each frame has, in
          addition, an infinite number of parents, one for each $U$.
          The frame fields have an inherent direction built in that
          is represented by stage labels as integers.

          Some properties of the frame fields are noted.  Each frame
          includes all physical theories as mathematical
          structures over the real and complex numbers
          irrespective of their use of a space time arena
          to describe systems dynamics. All frames at any one stage are
          unitarily equivalent and any two frames at the same or
          different stages are isomorphic to each other.

          The one way and two way infinite frame fields are
          different in that the base frame of a one way infinite
          field requires sets of real and complex
          numbers and a space time that is abstract and external in
          some absolute sense. They are not related to any frame.
          For the two way infinite frame field there are no external
          abstract real and complex numbers and no background  space
          time. All real and complex numbers and space
          times are external and abstract in a relative sense only in that they are
          external and abstract relative to a frame. However they
          are internal objects as sets of Cauchy operators in a
          parent frame.

          An important problem is to connect the field of frames to
          the one frame containing the observed physical universe.
          The goal of any process to achieve this is to
          identify as many frames as is possible with one frame.
          This is needed because it is not sufficient that frames in
          the frame field are isomorphic.  They must become the same
          in some limit. It is noted that the method of
          decoherence free subspaces \cite{Lidar,LidarII} can be
          used to identify all frames in a stage that are related
          by global gauge transformations.

          It is clear that much remains to be accomplished. However
          one has a very rich structure at hand with many
          possibilities to investigate.  Also the structure enables
          a new approach to physical problems in that the dynamics
          of qukit systems in a frame $F$ is reflected in the
          dynamics of the Cauchy operators in $F$. It is possible
          that this would be seen in a frame $F_{U}$ as some type
          of effect on the space time of the frame.  This suggests
          the possibility of a new way to approach the problem of
          combining gravity with quantum mechanics in that certain
          dynamics of quantum systems in $F$ appear in $F_{U}$ as
          the effect of gravity on physical systems.  It is also
          possible that the approach of Ashtekar and collaborators
          \cite{Ashtekar,AshtekarI,AshtekarII} to loop quantum
          gravity may be useful here.

          Other avenues to investigate include the use of
          different number bases by expanding gauge
          transformations  from those based on $SU(2)$ to
          those based on elements of $U(1)\times SU(2)\times
          \cdots \times SU(p_{m}).$ These are suitable for qukits
          where the base $k=p_{1}\cdots p_{m}.$
          Here $p_{m}$ is the $mth$ prime number.

          Other possible methods of frame identification include the
          possibility of turning the one and two infinite frame
          fields into cyclic frame fields by identifying two
          different stages. Also the possibility of using the least
          action principle in a Feynman path integral over the frame
          field needs to be examined.

          In any case it is pleasing to note that this approach of
          constructing iterated frame fields corresponds to
          definite steps towards  the much desired goal of
          constructing a coherent theory of physics and mathematics
          together \cite{BenTCTPM}.

          Finally it should be noted that the  structure of
          frames emanating from  frames has nothing to do with the
          Everett Wheeler view of multiple universes \cite{Everett}.
          If these multiple universes exist, then they would exist
          within each frame in the field.

           \section*{Acknowledgements}
       This work was supported by the U.S. Department of Energy,
       Office of Nuclear Physics, under Contract No. W-31-109-ENG-38.


\begin{thebibliography}{99}

          \bibitem{Aharonov}
          Y. Aharonov and T. Kaufherr, Phys. Rev. D \textbf{30},
          368-  (1984).

          \bibitem{Bagan}
          E. Bagan, M. Baig, and R. M$\tilde{u}$noz-Tapia,
          Phys. Rev. Lett. \textbf{87}, 167901, (2001).

          \bibitem{Rudolph}
          T. Rudolph and L. Grover, Phys. Rev. Lett. \textbf{91}, 217905,
          (2003).

          \bibitem{Bartlett}
          S. D. Bartlett, T. Rudolph, and R. W. Spekkens, Phys. Rev.
          A \textbf{70}, 032307 (2004).

          \bibitem{vanEnk}
          S. J. van Enk, Phys. Rev. A \textbf{71}, 032339 (2005).

           \bibitem{Enk}
          S.J. van Enk, Arxiv preprint quant-ph/0602079.

          \bibitem{BenRRCNQT}
          Paul Benioff, Arxiv preprint quant-ph/0508219.

          \bibitem{AharonovII}
          Yakir Aharonov and Leonard Susskind, Phys. Rev.
          \textbf{155},1428-1431, (1967)>

          \bibitem{Lidar}
          Mark S. Byrd, Daniel Lidar, Lian-Ao Wu, and Paolo Zanardi,
          Phys. Rev A \textbf{71}, 052301 (2005).

           \bibitem{BenTCTPM}
          Paul Benioff, Found. Phys. \textbf{35}, 1825-1856, (2005).

          \bibitem{Smolin}
          Lee Smolin, Arxiv preprint hep-th/0507235.

          \bibitem{Rovelli}
          Carlo Rovelli, Arxiv preprint gr-qc/0604045; \emph{Quantum Gravity},
          Cambridge University Press, Cambridge, UK, (2004).

          \bibitem{Horowitz}
          Gary Horowitz, ArXiv preprint gr-qc/0410049.

          \bibitem{BenRCRNQM}
          Paul Benioff, Phys. Rev. A \textbf{72}, 032314 (2005).

          \bibitem{Toller}
          M. Toller, Nuovo Cim. \textbf{B112}, 1013-1026, (1997),
          Arxiv preprint, gr-qc/9605052.

          \bibitem{Takeuti}
          Gaisi Takeuti, \emph{Two Applications of Logic to
          Mathematics} Kano Memorial Lecture 3, Princeton University
          Press, New Jersey, 1978.

          \bibitem{Davis}
          Martin Davis, Internat. Jour. Theoret. Phys.
          \textbf{16},867-874,(1977).

          \bibitem{Gordon}
          E. I. Gordon, Soviet Math. Dokl. \textbf{18}, 1481-1484
          (1977).

            \bibitem{Litvinov}
            G. L. Litvinov, V. P. Maslov, and G. B. Shpiz,
            Archives preprint, quant-ph/9904025, v5, 2002.


            \bibitem{Corbett}
            J. V. Corbett and T. Durt, Archives preprint,
            quant-ph/0211180 v1 2002.

            \bibitem{Tokuo}
            K. Tokuo, \emph{Int. Jour. Theoretical Phys.},
            \textbf{43}, 2461-2481, 2004.


          \bibitem{Krol}
          Jerzey Krol, "A Model of Spacetime. The Role of
          Interpretations in Some Grothendieck Topoi", preprint,
          (2006).

          \bibitem{Ng}
          Y. Jack Ng and H. Van Dam, Int. Jour. Mod. Phys. A
          \textbf{20}, 1328-1335 (2005).

          \bibitem{Girelli}
          Florian Girelli and Etera R. Levine Arxiv preprint
          gr-qc/0501075.

          \bibitem{Terno}
          Damiel Terno, Arxiv preprint gr-qc/0512072.

          \bibitem{Perez}
          Alejandro Perez, Arxiv preprints,  gr-qc/0601095; gr-qc/0409061.

          \bibitem{Novaes}
          S. F. Novaes, Arxiv preprint hep-th/0001283.

          \bibitem{Cottingham}
          A. N. Cottingham and D. A. Greenwood, \emph{An Introduction
          to the Standard Model of Physics}, Cambridge University
          Press, Cambridge, UK, 1998.


          \bibitem{LidarII}
          J. Kempe, D. Bacon, D. A. Lidar, and K. B. Whaley, Phys.
          Rev. A \textbf{63}, 042307 (2001).

          \bibitem{Ashtekar}
          Abhay Ashtekar, New Journal of Physics, \textbf{7}, 198
          (2005), Arxiv preprint, gr-qc/0410054.

          \bibitem{AshtekarI}
          Abhay Ashtekar, Arxiv preprint, gr-qc/99010123.

          \bibitem{AshtekarII}
          Abhay Ashtekar and Jerzy Lewandowski, Arxiv preprint,
          hep-th/9603083.

          \bibitem{Everett}
          Hugh Everett III,  Rev. Mod. Phys. \textbf{29},
          454-462, (1957); John A. Wheeler, Rev. Mod. Phys.
          \textbf{29}, 463-465 (1957).

          \end{thebibliography}
          \end{document}